\documentclass[12pt, draftclsnofoot, onecolumn]{IEEEtran}
\usepackage{subfig}
\usepackage{float}
\usepackage{amssymb}

\usepackage{lipsum}

\usepackage{amsmath}
\usepackage{graphicx}
\usepackage{comment}
\usepackage{cite}
\DeclareMathOperator*{\argmax}{arg\,max}
\DeclareMathOperator*{\argmin}{arg\,min}

\title{\LARGE \bf
A New Codebook Design for Analog Beamforming in Millimeter-wave Communication
}

\begin{document}

\author{Mehdi~Ganji,~\IEEEmembership{Student Member,~IEEE,}
Hongbing~Cheng,~\IEEEmembership{Senior Member,~IEEE,}
Qi~Zhan,~\IEEEmembership{Member,~IEEE,}
Kee-Bong~Song,~\IEEEmembership{Member,~IEEE}\thanks{M. Ganji is with the Center for Pervasive Communications and Computing, University of California, Irvine, CA, 92697 USA (e-mail: mganji@uci.edu). H. Cheng, Q. Zhan and K. Song are with SoC Lab, Samsung Semiconductor Inc., San Diego, CA, 92121 USA (e-mail: \{hongbing.c\},\{qi.zhan\},\{keebong.s\}@samsung.com). This work was conducted while the first author was an engineering intern at Samsung Semiconductor Inc., San Diego, CA.}}

\maketitle

%%%%%%%%%%%%%%%%%%%%%%%%%%%%%%%%%%%%%%%%%%%%%%%%%%%%%%%%%%%%%%%%%%%%%%%%%%%%%%%%
\begin{abstract}
In this study, we analyze the codebook design used for analog beamforming. Analog beamforming and combining suffer from a subspace sampling limitation, that is, the receiver cannot directly observe the channel coefficients; instead, the receiver observes a noisy version of their weighted combination. To resolve this, the transmitter and the receiver usually collaborate to determine the best beamformer combiner pair during the beam-sweeping process. This is done by evaluating a limited number of codewords chosen from a pre-defined codebook. In this study, we propose a new framework inspired by the generalized Lloyd algorithm to design analog beamforming codebooks that optimize various performance metrics including the average beamforming gain, the outage, and the average data rate. The flexibility of our framework enables us to design beamforming codebooks for any array shapes including uniform linear and planar arrays. The other practical complexity in analog beamforming is the low resolution of the phase shifters. Therefore, we have extended our algorithm to create quantized codebooks that outperform the existing codebooks in literature. We have also provided extensive simulations to verify the superiority of our proposed codebook as compared with the existing codebooks.
\end{abstract}

%%%%%%%%%%%%%%%%%%%%%%%%%%%%%%%%%%%%%%%%%%%%%%%%%%%%%%%%%%%%%%%%%%%%%%%%%%%%%%%%
\section{Introduction}
Millimeter-wave (mmWave) communication is a promising technology for next-generation wireless communication because of its abundant frequency spectrum resource, which promises a much higher capacity than the current cellular mobile communication. In the mmWave domain, mixed signal components consume very high power and radio frequency (RF) chains are expensive. This hinders the realization of digital baseband beamforming as in conventional multiple-input multiple-output (MIMO) systems \cite{jafarkhani2005space,gershman2005space,duman2008coding,ganji2016interference,yu2016alternating} particularly in Massive MIMO settings \cite{larsson2014massive,lu2014overview,ganji2018performance, bjornson2016massive}. Therefore, analog beamforming is usually preferred when all the antennas share a single RF chain and have constant-amplitude constraints on their weights \cite{kutty2016beamforming}. 

With one RF chain, the channel coefficients cannot be directly observed. Instead, a noisy version of their linear combination weighted by a vector called the beamforming vector (codeword) is accessible. In the beam-sweeping process, the beamforming vectors are chosen from a pre-defined pool (codebook) of antenna weight vectors that are maintained at the transmitter and the receiver \cite{wang2009beam}. After the beam-sweeping process, we can attempt to estimate the channel coefficients and use the estimated channel to perform beamforming. Another method is to choose the beamformer/combiner pair that optimizes a certain cost function and use that pair for subsequent transmissions. We call this method as selection-based beamforming, and this will be the focus of our study.

The constant modulus constraint imposed by the phase shifters is a major obstacle to optimizing the beamforming designs \cite{el2014spatially}. One common solution is to consider the discrete Fourier transform (DFT) codebook, which includes $N$ beams, $N$ being the number of antennas in the array. However, the DFT beams have a narrow beam width and limited sweeping resources; therefore, they provide limited coverage and performance. Hence, the challenge is to design a codebook of beamforming vectors with wide beam widths \cite{hur2013millimeter}. In addition, to improve the search efficiency, we may define a hierarchical of codebooks \cite{chen2011multi,alkhateeb2014channel,chen2017efficient}. For example, a coarse codebook may be defined by using a small number of low-resolution beams that cover the intended angle range, whereas a fine codebook may be defined with a large number of high-resolution beams that cover the same intended angle range. In \cite{xiao2016hierarchical}, the authors proposed an approach called BeaM Widening via SingleRF Subarray (BMW-SS), where they jointly used the subarray and the deactivation approach to create wide beams. To create wide beams, however, half of the array antennas might need to be deactivated; this limits the application of BMW-SS for mmWave communication. In addition, the proposed approach may not be extended to arrays with an arbitrary number of antenna elements. In \cite{wang2009beam}, a three-level hierarchical codebook was proposed, but it provided a limited beamforming gain. In \cite{chen2011multi}, codewords with wide beams were generated by summing two codewords with narrow beams, but the constant module constraint was not satisfied. In \cite{alkhateeb2014channel,pezeshkpour2014optimal}, the hybrid precoding structure was used to shape wide beams by exploiting the sparse reconstruction approach, but the wide beams could be shaped accurately only when the number of RF chains was large enough; otherwise, deep sinks appeared within the spatial domain. 

Another method for creating wider beams falls under the beam-pattern synthesis (BPS) problem \cite{fuchs2014application,liang2017phase}, where the goal is to design the beamforming vector (codeword) such that it generates some pre-defined beam patterns. For example, in \cite{raviteja2017analog}, the preferred beamforming vectors were selected to approximate a beamforming gain mask by using a low complexity local search algorithm. In \cite{song2015codebook}, the analog codewords were designed to minimize the MSE between the pre-defined beam patterns and the beam pattern generated by the beamforming vector. In \cite{zhang2017codebook}, the authors proposed a codebook design by formulating the BPS as an optimization problem by using both the ripple and the transition band of beam patterns.

However, imposing pre-defined shapes on the beam patterns limits both the performance of the codebook and its adaptability to different array shapes, such as Uniform Linear Array (ULA), Uniform Planar Array (UPA) and Uniform Circular Array (UCA). In addition, considering a hierarchical design is mainly beneficial when the channel is modeled as a one-ray channel (Line-of-sight); otherwise, the hierarchical design might not capture all the available power \cite{xiao2016hierarchical}. Also, there is error propagation in the search process, which can reduce the beamforming performance. The design of the analog-only beamforming vectors with certain shapes or beam widths relies mostly on the beam steering of ULAs. It is difficult to apply this design to non-ULAs because of the lack of intuition about their beam patterns \cite{choi2015advanced,song2017common}. Indeed, most antennas already deployed in cellular systems are UPAs, which necessitates the need for a more flexible framework considering different antenna configurations. Besides, the existence of quantized phase shifters complicates the design of non-overlapping beam patterns, which might require an exhaustive search over a large space given the large number of antennas \cite{alkhateeb2014channel}. However, the search complexity involved in determining the optimal beamformers by using the exhaustive search algorithm increases exponentially with the number of antennas and the resolution of the phase shifters; this limits and complicates their implementation \cite{chen2017efficient,raviteja2017analog}. Quantized beamforming with perfect knowledge of channel state information (CSI) is also considered in \cite{wang2018hybrid,seleem2017hybrid}. However, because of the large size of the antenna arrays and the limited number of sweeping resources, the acquisition of CSI matrices could be challenging and would require a more efficient design for the quantized analog beamforming vectors for UPAs without any CSI information \cite{javani2019age1,javani2019age2}. In addition, computational complexity and timing play a central role in the algorithm design \cite{ pourghassemi2017cudacr,mostafazadeh2017unsteady,pourghassemi2019if,shahhosseini2019dynamic,shahosseini2017dependability}.

In this paper, we propose a new framework to design analog beamforming codebooks that enable us to design analog beam vectors without any restrictions on the beam patterns, the number of antennas, or the antenna array shape. We model the codebook design for analog beamforming as a vector quantization problem \cite{gray1984vector} to seek the codebook
that optimizes the performance metric over all possible codebooks. A widely used design approach defines the data source through the use of a finite training set, and uses an iterative procedure called the generalized Lloyd algorithm to generate a codebook \cite{gersho2012vector}. Similarly, our iterative method uses training channel vectors generated based on the system characteristics to output a locally optimal codebook. After the initialization step, the iteration begins by assigning each training channel vector to its ``best fit'' codeword based on some closeness (distortion) measure and an exhaustive search. Next, given a set of training vectors assigned to a particular codeword, the corresponding codeword is modified to optimize the performance metric relative to its currently assigned training vectors. The algorithm is essentially the same as the k-means algorithm used in pattern recognition \cite{jain1999data} where the codebook design can be interpreted as an unsupervised learning problem to optimally cluster all the training channel vectors and find an optimal codeword for each cluster in order to optimize a desired metric. Unlike \cite{alkhateeb2016frequency}, where an algorithm was proposed to minimize a specific distortion function for ULAs, our generalized Lloyd-based (LB) framework can be applied to various performance metrics including the average beamforming gain, outage, and the average data rate. In addition, we applied the proposed algorithm to the UPAs that were widely used in the 3GPP standards (see 3GPP TR 38.901) with an arbitrary number of antennas. Due to the flexibility of our proposed method even with a low-resolution phase shifter, this method can outperform existing analog beamforming methods with unlimited resolutions.

In summary, the contributions of this research are as follows:
\begin{itemize}
\item We propose a general LB algorithm to design analog beamforming codebooks using different number of codewords. 
\item We consider different antenna array shapes including ULA, which is the most common array shape in the literature, and UPA, which is more practical and insufficiently investigated in literature.
\item We propose various codebooks intended for optimizing different metrics including the average beamforming gain, the average data rate, and the outage. 
\item We also consider quantized phase shifters and show that our proposed codebooks outperform the existing codebooks having ideal phase shifters. 
\item We compare the performance of our proposed LB codebook with the DFT (beam-steering), BMW-SS \cite{xiao2016hierarchical}, and BPS \cite{liang2017phase} codebooks using extensive simulations. 
\end{itemize}
The rest of paper is organized as follows.
In Section \ref{model}, we present the general settings, including the spatial response description in the 3GPP antenna model, the system model assumptions, and the evaluation metrics. In Section \ref{litreture}, we present some codebooks in literature and explain their drawbacks. In Section \ref{framework}, we introduce our proposed framework including some preliminaries on the Lloyd's algorithm and our generalized algorithm for designing codebooks. We also consider the quantized phase shifters. Then, we provide extensive performance simulations in Section \ref{simulation}. Finally, we present our concluding remarks in Section \ref{conclusion}.
\vspace{-10pt}
\section{General Settings}\label{model}
We consider a mmWave communication
system with antenna arrays of $N_T$ and $N_R$ elements equipped at the transmitter and
receiver, respectively. Please note that we assumed single antenna array and single RF chain throughout the report, however, the codebook designs can be extended to multiple antenna arrays and multiple RF chains \cite{xiao2017codebook}. 
\vspace{-10pt}
\subsection{Spatial Response}
\begin{figure}[h]
    \centering
    \includegraphics[width=2.5in]{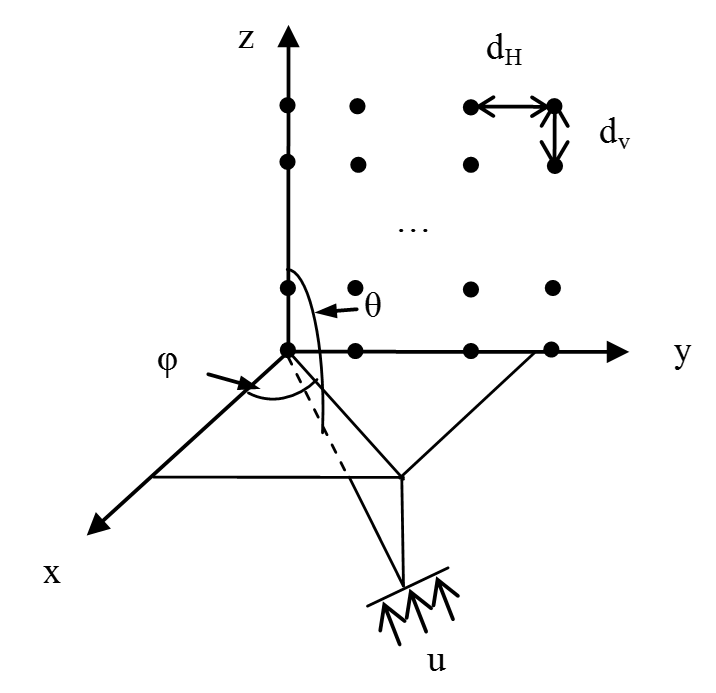}
    \caption{3GPP coordinate system for antenna array }
    \label{coordinate}
\end{figure}

We first describe 3GPP antenna model [e.g. see TR 38.901]. Antennas are defined in 3D coordinate system as in Fig. \ref{coordinate}. The
incoming ray is represented by a pair of two angles $(\theta,\phi)$, where $\theta$ denotes angle to the zenith and $\phi$ is the angle between the projection of incoming ray to x-y plane and the x axis. Antennas are located in y-z plane with $N_v$ elements in vertical direction and $N_h$ elements in horizontal direction. The spacing between two elements are $d_v$ and $d_h$ in vertical and horizontal directions, respectively.
For 1D array, $N_h$ will reduce to 1 and array response in $\phi$ is identical. We define antenna index $(n_v,n_h)$ as the $n_v$th antenna in vertical column and the $n_h$th antenna in horizontal row. Assuming a planar wave model, the location of antenna elements introduce phase offsets. The offset at Antenna $(n_v,n_h)$ can be written as \cite{stutzman2013antenna}:
\begin{flalign}
\nonumber
v_{n_v,n_h}(\theta,\phi)=&exp\left(j2\pi\left((n_v-1)\frac{d_v}{\lambda}cos(\theta)+(n_h-1)\frac{d_h}{\lambda}sin(\theta)sin(\phi) \right)\right)\\
&\ \ \ \ \ \ n_h=1,\cdots,N_h, \ \ n_v=1,\cdots,N_v
\label{offset}
\end{flalign}
Assume that Antenna $(n_v,n_h)$ applies phase shift of $w_{n_v,n_h}$, then the array response of antenna system is defined as
$A(w,\theta,\phi)=\left|\sum_{n_v,n_h}{w^*_{n_v,n_h}v_{n_v,n_h}(\theta,\phi)}\right|$.
The spatial response or the beam pattern of the antenna system is then defined as $B(w,\theta,\phi)=A(w,\theta,\phi)P_E(\theta,\phi)$, where $P_E(\theta,\phi)$ denotes the directionality of antenna elements and for an isotropic element $P_E(\theta,\phi)=1$. In this work, we consider isotropic elements, however, our proposed framework can also be applied to directional antenna elements. For simple and unified representation, we define $N_v\times N_h$ matrices of $\boldsymbol{W}$ and $\boldsymbol{V}(\theta,\phi)$ containing $w_{n_v,n_h}$ and $v_{n_v,n_h}(\theta,\phi)$ as their $(n_v,n_h)$th element, respectively. Then, the spatial response (beam pattern) of the corresponding codeword can be written as
$B(\boldsymbol{w},\theta,\phi)=\left|\boldsymbol{w}^H\boldsymbol{v}(\theta,\phi)\right|^2$
where $\boldsymbol{w}=vec(\boldsymbol{W})$ and $\boldsymbol{v}(\theta,\phi)=vec(\boldsymbol{V}(\theta,\phi))$. Note that the mentioned representation can constitute any antenna array shape with $N$ antenna elements given the corresponding phase offsets vector. 
\vspace{-10pt}
\subsection{System Model}
The received signal at the receiver can be written as: 
\begin{flalign}
y=\sqrt{P_{tot}}\boldsymbol{w}_{\boldsymbol{R}}^H\boldsymbol{H}\boldsymbol{w_T}s+\boldsymbol{w}_{\boldsymbol{R}}^H\boldsymbol{n}
\end{flalign}
where $s$ denote the transmitted symbol with unit power, $\boldsymbol{w_T}$ and $\boldsymbol{w_R}$ are the $N_T\times 1$ transmit and $N_R\times 1$ receive beamforming vectors, respectively, $\boldsymbol{H}$ is the $N_R\times N_T$ channel matrix and $\boldsymbol{n}$ is the Gaussian noise
vector with power $N_0$, i.e., $E(\boldsymbol{n}\boldsymbol{n}^H) = N_0\boldsymbol{I_{N_R}}$. Beamforming vectors are chosen from $
\mathit{W}(N)=\left\{\left.\frac{1}{\sqrt{N}}[e^{j\theta_1},e^{j\theta_2},\cdots,e^{j\theta_N}]\right|\theta_i\in [0,2\pi), i=1,\cdots,N \right\}
\label{set}$
where $\frac{1}{\sqrt{N}}$ is a normalization factor such that all the vectors have unit power \cite{xiao2016hierarchical}.

The channel matrix can be modeled
by the combination of a line-of-sight (LOS) path and a few non-line-of-sight (NLOS) paths as \cite{hsu2015low}:
\begin{flalign}
\boldsymbol{H}=\sqrt{\frac{\kappa}{\kappa+1}}\boldsymbol{v_{R}}(\theta^0_R,\phi^0_R)\boldsymbol{v}_{\boldsymbol{T}}^H(\theta^0_T,\phi^0_T)+\sqrt{\frac{1}{I(\kappa+1)}}\sum_{i=1}^{I}\alpha_i\boldsymbol{v_{R}}(\theta^i_R,\phi^i_R)\boldsymbol{v}_{\boldsymbol{T}}^H(\theta^i_T,\phi^i_T)
\end{flalign}
where $\kappa$ is the Ricean K-factor, $\alpha_i \sim CN(0,1)$ is the
complex channel gain, $I$ is the number of NLOS paths. Each  ray is represented by a pair of two angles $(\theta,\phi)$ where $\theta$ denotes angle to the zenith and $\phi$ is the angle between the projection of incoming ray to x-y plane and the x axis as shown in Fig. \ref{coordinate}. $\boldsymbol{v_{\{R/T\}}}(.,.)$ is a $N_{\{R/T\}}\times 1$ vector representing the phase offsets introduced at receive/transmit antenna elements. 
The average power transmitted from each transmit antenna equals $P_{tot}/N_{T}$ and the total radiated power from the transmitter is thus equal to $P_{tot}$. After combining the signals from all receive antenna elements, the effective received signal-to-noise ratio (SNR) is defined as $\delta_{R}=P_{tot}|\boldsymbol{w}_{\boldsymbol{R}}^H\boldsymbol{H}\boldsymbol{w_T}|^2/N_0$. The total beamforming gain due to beamforming at the transmitter and the receiver can be denoted as $G_B=|\boldsymbol{w}_{\boldsymbol{R}}^H\boldsymbol{H}\boldsymbol{w_T}|^2$ \cite{xiao2016hierarchical}. 
\vspace{-10pt}
\subsection{Selection-based Beamforming}
In the selection-based beamforming, a set of codewords is chosen from $\mathit{W}(N)$ to create the beamforming codebook. After evaluating all the codewords within the pre-designed codebook, the \textit{best} one which maximizes the desired metric is chosen and used in successive transmissions. Assuming $K$ sweeping resources, then, the beamforming codebook is denoted as an $N\times K$ matrix, $\mathbb{W}$, containing beamforming codewords on its columns. Then, the commonly used beam-sweeping process refereed as \textit{ping-pong sampling} \cite{hur2013millimeter} can be explained as follows. During Tx beam-sweeping, the receive beamforming vector can be set to omni-directional beam and then the obtained sample by evaluating the $k$th codeword in the Tx codebook can be written as:
\begin{flalign}
y[k]=\sqrt{P_{tot}}{\boldsymbol{w}_{\boldsymbol{R}}^{omni}}^H\boldsymbol{H}\mathbb{W_T}[k]+{\boldsymbol{w}_{\boldsymbol{R}}^{omni}}^H\boldsymbol{n}[k], \ \ \ k=1,\cdots, K
\end{flalign} 
where $K$ is the number of Tx beam-sweeps. The receiver feedbacks the best Tx beamforming vector, i.e.,
$\boldsymbol{w}_{\boldsymbol{T}}^{opt}=\mathbb{W_T}[k_{opt}], \ \ \text{with} \ \ \ k_{opt}=\argmax_k{|y[k]|^2}$.
After Rx measurement and feedback, the Tx beam is selected and will be kept the same in the Rx beam-sweeping process. At the receiver side, all the codewords in the Rx codebook are applied. Similarly, the receiver chooses the best receive beamforming vector, i.e., $\boldsymbol{w_R}^{opt}$ by observing $K'$ samples of 
\begin{flalign}
y[k']=\sqrt{P_{tot}}{\mathbb{W_R}[k']}^H\boldsymbol{H}\boldsymbol{w}_{\boldsymbol{T}}^{opt}+\mathbb{W_R}[k']^H\boldsymbol{n}[k'], \ \ \ \ k'=1,\cdots,K'
\end{flalign} 
where $K'$ is the number of Rx beam-sweeps. Then, the pair of $(\boldsymbol{w}_{\boldsymbol{T}}^{opt},\boldsymbol{w}_{\boldsymbol{R}}^{opt})$ is used for successive transmissions. Consequently, the effective beamforming gain associated with Tx and Rx codebooks can be denoted as  $G_B(\mathbb{W_R},\mathbb{W_T})=|{\boldsymbol{w}^{opt}_{\boldsymbol{R}}}^H\boldsymbol{H}\boldsymbol{w}^{opt}_{\boldsymbol{T}}|^2$. Because the beamforming processes are the same at the transmitter and the receiver, we focus on designing the codebook at the receiver side and the same codebook can be used at the transmitter (the index of T/R is discarded). 
\vspace{-10pt}
\subsection{Metrics for Designing Codebooks }
\subsubsection{Average beamforming gain}
Different metrics can be considered to compare the codebooks. One of the common metrics is the average of beamforming gain over different channel realizations, defined as $\overline{G_B}(\mathbb{W})=E_{\boldsymbol{h}}\left[G_B(\mathbb{W})\right]$
where $G_B(\mathbb{W})=\left|{\boldsymbol{w}^{opt}}^H\boldsymbol{h}\right|^2$. Neglecting the beam mis-detection, the effective beamforming gain reduces to $G_B(\mathbb{W})=\max_{k}\left|{\mathbb{W}[k]^H\boldsymbol{h}}\right|^2$ and hence $\overline{G_B}(\mathbb{W})$ can be written as $E_{\boldsymbol{h}}\left[\max_{k}\left|{\mathbb{W}[k]^H\boldsymbol{h}}\right|^2\right]$. Then, the codebook design can be formulated as designing the matrix $\mathbb{W}$ which includes $K$ codewords from the feasible set $W(N)$ maximizing:
\begin{flalign}
J_{avg}=E_{\boldsymbol{h}}\left[\max_{k=1,\cdots,K}\left|{\mathbb{W}[k]^H\boldsymbol{h}}\right|^2\right]
\label{metric}
\end{flalign}
\subsubsection{Outage}
Considering the randomness in the channel vectors, the effective beamforming gain, i.e., $G_B(\mathbb{W})$, is a random variable whose distribution depends on distribution of $\boldsymbol{h}$ and also the beamforming codebook. Although the average metric, i.e., $\overline{G_B}(\mathbb{W})$, is an important metric in analyzing system performance, in some applications, the individual performance in each realization is of greater importance. This motivates us to define the outage metric as:
\begin{flalign}
J_{out}(\gamma)= Pr\left\{ \max_{k=1,\cdots,K}\left|{\mathbb{W}[k]^H\boldsymbol{h}}\right|^2 < \gamma \right\}
\label{out}
\end{flalign}
where $\gamma$ is a performance threshold specified by the system requirements. The coverage can be defined as the complimentary event of outage as $J_{cov}(\gamma)=1-J_{out}(\gamma)$. Normalizing the threshold by the maximum possible gain, i.e., $\hat\gamma=\gamma/N$, $J_{cov}(\gamma)$ and $J_{out}(\gamma)$ can be called the $\hat\gamma$-coverage and the $\hat\gamma$-outage, respectively. For example, for a $2\times2$ UPA, $J_{cov}(2)$ and $J_{cov}(1)$ are called $1/2$-coverage and $1/4$-coverage or conventionally 3dB-coverage and 6dB-coverage, respectfully.

To have a unified framework, we rewrite the outage metric as follows $
J_{cov}(\gamma)= E_{\boldsymbol{h}}\left[sign\left(\max_{k}\left|{\mathbb{W}[k]^H\boldsymbol{h}}\right|^2-\gamma\right)\right]$
where $sign(x)=\left\{\begin{matrix}
1 \ \   x\geq 0\\ 
0  \ \   x<0
\end{matrix}\right.$. We further simplify the outage metric by replacing the $sign(\dot)$ function which is not differentiable with $sigmoid$ function defined as: $sigmoid(x)=\frac{1}{1+e^{-\alpha x}}$ where $\alpha$ is a metric to adjust the steepness of the curve. Thus, the coverage metric can be approximated by $J_{cov}(\gamma)\approx E_{\boldsymbol{h}}\left[sigmoid\left(\max_{k}\left|{\mathbb{W}[k]^H\boldsymbol{h}}\right|^2-\gamma\right)\right]$.
\subsubsection{Average Data Rate}
Average data rate is another important criterion which can be defined as:
\begin{flalign}
J_{rate}=E_{\boldsymbol{h}}\left[log_2\left(1+\max_{k=1,\cdots,K}\left|{\mathbb{W}[k]^H\boldsymbol{h}}\right|^2\right)\right]
\label{metric}
\end{flalign}
Other metrics, e.g. bit error rate (BER) performance, can also be considered, thus we propose a unifying approach that can be applied to various performance metrics. Before delving into details of our proposed framework, some of the codebooks in the literature are analyzed in the next Section.
\vspace{-10pt}
\section{Problem statement}\label{litreture}
\begin{figure}[h]
    \centering
    \includegraphics[width=2.5in]{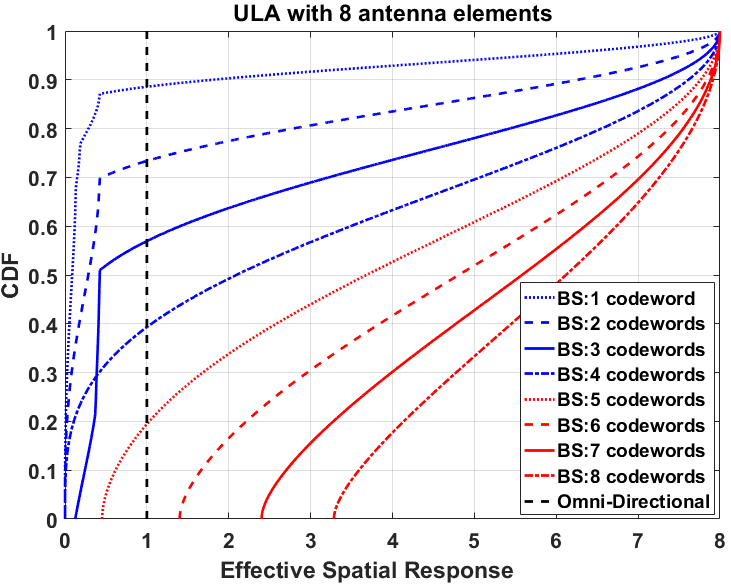}
    \caption{CDF of effective spatial response}
    \label{dft}
\end{figure}

Here, we present and discuss some of the existing beamforming codebooks and their drawbacks. Most of the work in the corresponding literature considers the 1D antenna array (ULA), for which the steering vector reduces to a vector, the only input of which is the angle to the zenith, $\boldsymbol{v}(\theta) = [1,e^{j2\pi\frac{d}{\lambda}cos(\theta)},\cdots,e^{j(N-1)2\pi\frac{d}{\lambda}cos(\theta)}]^T$. Assuming a uniform distribution for $\theta $ over the spatial domain $\Psi$, the spatial response can be considered a random variable for any given codeword. For example, for the omni-directional codeword, the spatial response is equal to 1, i.e., $B(\boldsymbol{w})=1$, with a probability of 1. The other common codeword is termed the beam-steering codeword and is defined as $\boldsymbol{w}=\frac{1}{\sqrt{N}}\boldsymbol{v}(\theta)$. The beam-steering codeword can achieve the largest array gain toward its directionality, i.e., $N$; however, owing to its narrow beam, it has poor performance for other values of $\theta$. To further analyze the behavior of a codebook, we extend the definition of spatial response to $B(\mathbb{W})=\max_{k}\left|{\mathbb{W}[k]^H\boldsymbol{v}}\right|^2$, which is termed the effective spatial response of the codebook. Assuming $\theta$ to be uniformly distributed in $[0,180]$, the cumulative density function (CDF) of the effective spatial response for the omni-directional codeword and beam-steering codebook with different numbers of codewords is shown in Fig. \ref{dft}.
The beam width of the Beam-Steering codewords are inversely proportional to the number of antennas, meaning that, when we increase the number of antennas the beams get narrower with smaller beam width. Therefore, with limited number of codewords, the beam-steering codebook fails to provide reliable performance as shown in Fig. \ref{dft}. It can be observed that 1,2, or even 3 beam-steering codewords provide a worse performance than omni-directional codeword with the probability of more than half. By increasing the number of codewords, more reliable performance is achieved, specifically, for DFT codebook with $N=8, Pr(B(\mathbb{W})<3.2)=0$. 

The main problem in analog beamforming is that the number of beam-sweeping resources is usually much lower than that of antenna elements. Therefore, the DFT codebook is not a good candidate owing to narrow beam width of its codewords. One solution offered in literature is deactivation, whereby the antenna elements are deactivated to provide wider beams \cite{he2015suboptimal}. However, to create wide beams, the number of active antennas becomes small, which limits the maximal total transmission/reception power of a mmWave device. Another solution is using the sub-array technique. In particular, if a large antenna array is divided into multiple sub-arrays and these sub-arrays point at sufficiently-spaced directions, a wider beam can be shaped. The combination of these two methods is termed BMW-SS in \cite{xiao2016hierarchical} and follows a hierarchical design with $log_2(N)$ levels, with $N$ being the number of antenna elements. 
\begin{figure}[!h]
\centering
\subfloat[level 1: 2 codewords]{ 
\includegraphics[width=2in]{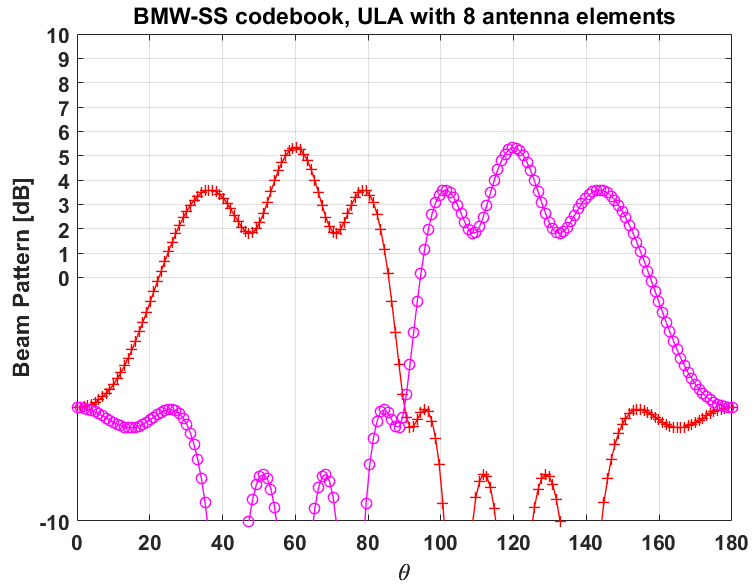}
}
\hfil
\subfloat[level 2: 4 codewords]{
\includegraphics[width=2in]{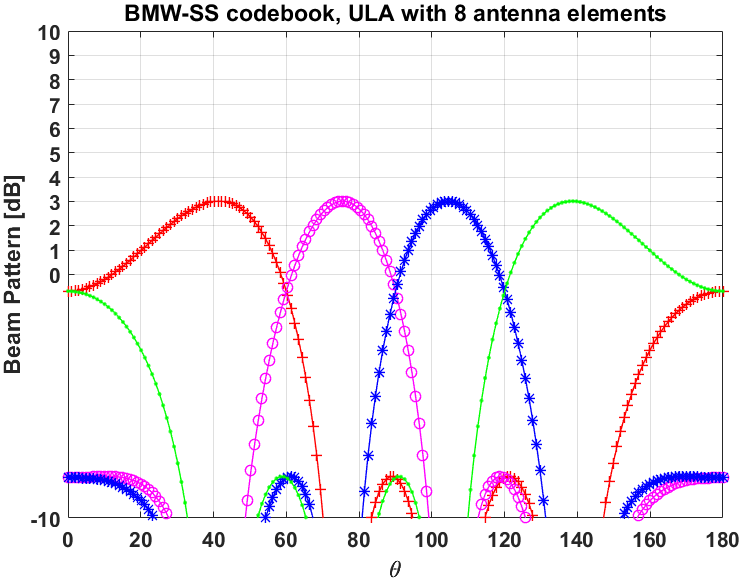}
}
\hfil
\subfloat[level 3: 8 codewords]{
\includegraphics[width=2in]{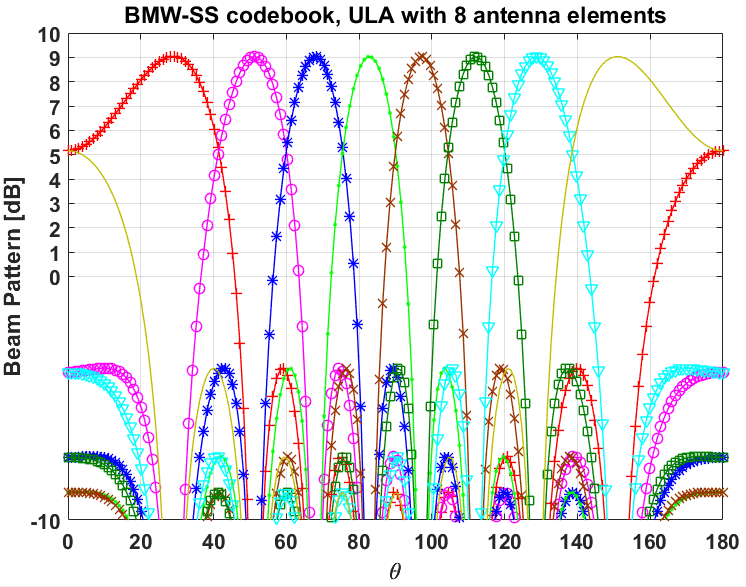}
}
\caption{Beam patterns of BMW-SS codebook}
\label{bmw-ss}
\end{figure}
For example, BMW-SS codebook for ULA with 8 antenna elements is shown in Fig. \ref{bmw-ss}. Note that the last level coincide with the DFT codebook. It can be seen that, the codewords in the second level suffer from power reduction due to antenna deactivation. In addition, BMW-SS codebook is only applicable to ULAs with the number of elements being a power of two.  

\begin{figure}[!h]
\centering
\subfloat[level 1: 2 codewords]{ 
\includegraphics[width=2in]{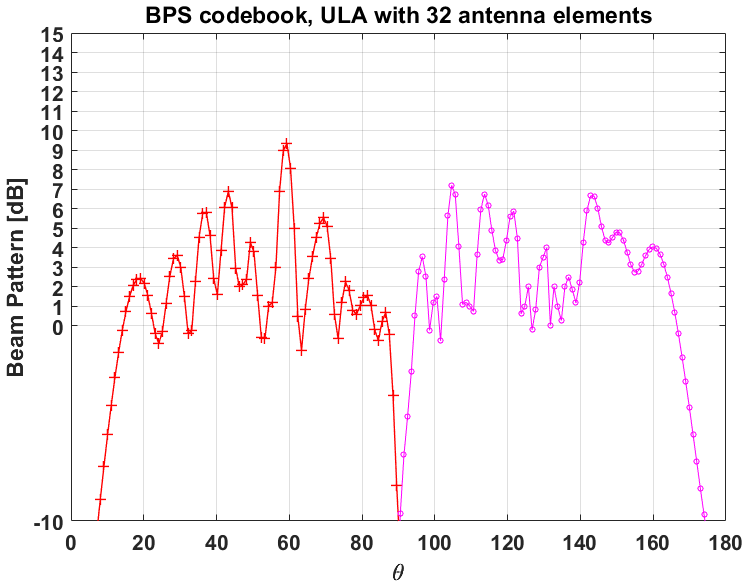}
}
\hfil
\subfloat[level 2: 4 codewords]{
\includegraphics[width=2in]{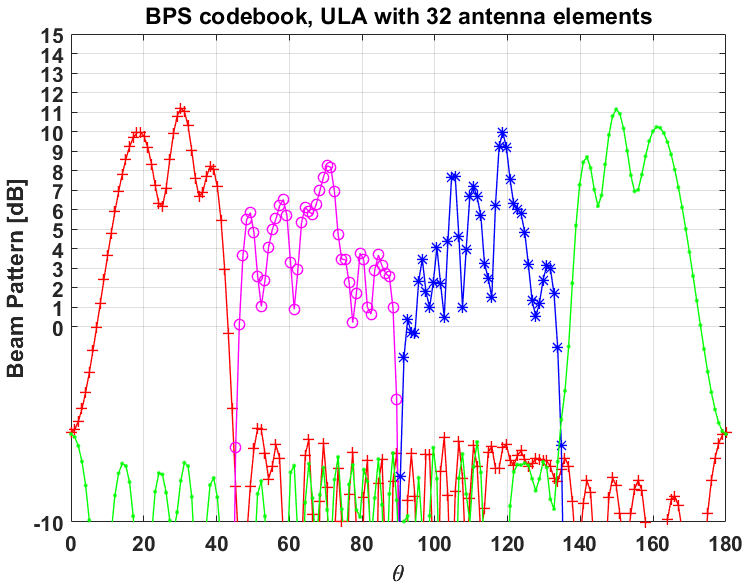}
}
\hfil
\subfloat[level 3: 8 codewords]{
\includegraphics[width=2in]{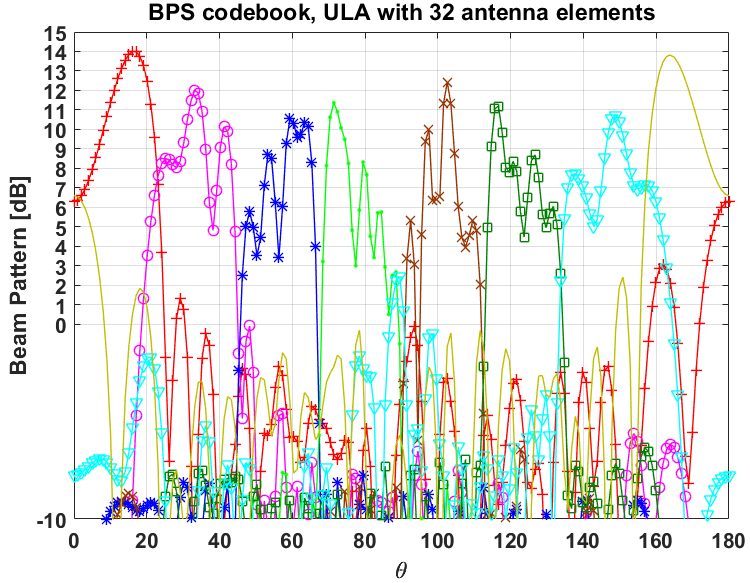}
}
\caption{Beam patterns of BPS codebook}
\label{ps}
\end{figure}

\begin{figure}[!h]
\centering
\subfloat[3dB-coverage]{ 
\includegraphics[width=2in]{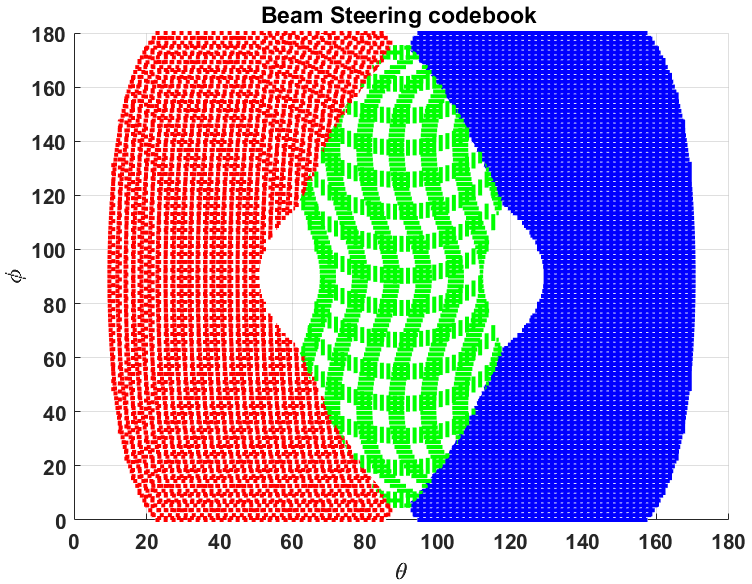}
}
\hfil
\subfloat[3D beam patterns]{
\includegraphics[width=2in]{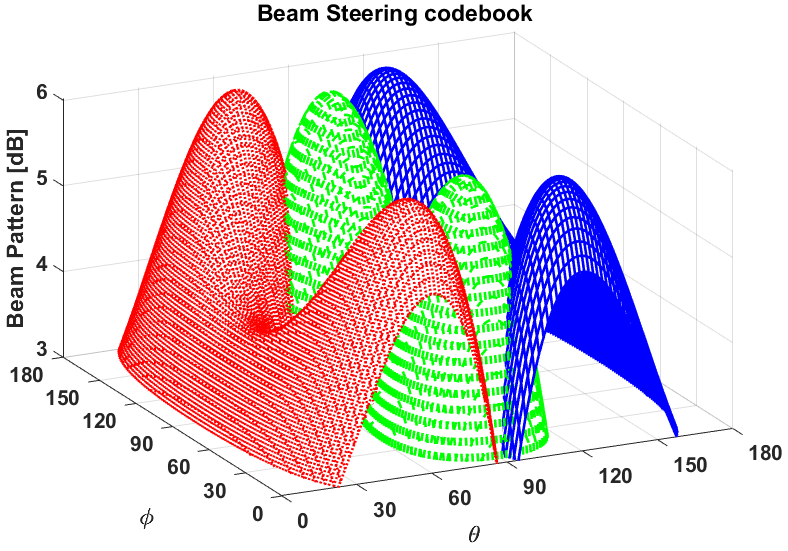}
}
\caption{Beam patterns of beam-steering codebook with 3 codewords}
\label{ps3}
\end{figure}

\begin{figure}[!h]
\centering
\subfloat[3dB-coverage]{ 
\includegraphics[width=2in]{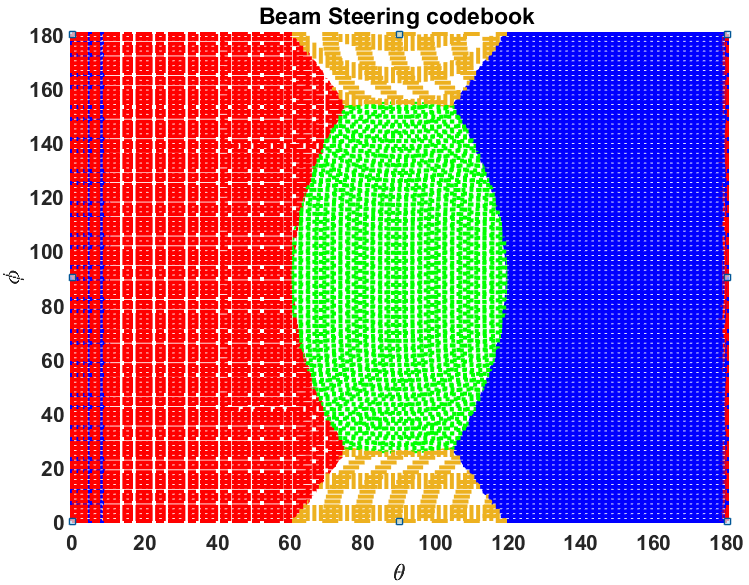}
}
\hfil
\subfloat[3D beam patterns]{
\includegraphics[width=2in]{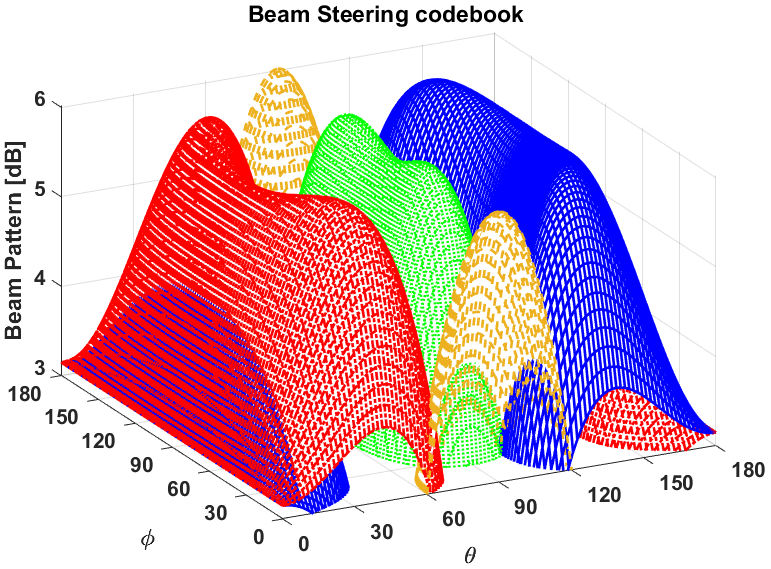}
}
\caption{Beam patterns of beam-steering codebook with 4 codewords}
\label{ps4}
\end{figure}

There are other works in the literature which constraint the spatial response to some pre-defined shapes and then attempt to synthesize it with various algorithms \cite{raviteja2017analog,zhang2017codebook}. For example, in Fig. \ref{ps}, the algorithm in \cite{liang2017phase} is used to generate a codebook with three levels for ULA with 32 antenna elements. Imposing pre-defined shapes on beam patterns, e.g., rectangular-like beam patterns, limits the performance and requires large number of antennas to be realizable \cite{song2015codebook}.  
Unlike ULA where many codebooks are proposed in the literature, there are very few works considering 2D antenna arrays, in particular UPA. Authors in \cite{xiao2016hierarchical} propose using Kronecker product to extend the 1D codebooks to 2D arrays. Similarly, beam pattern synthesis methods are proposed to synthesize the cubic-like beam patterns \cite{song2017common}. 
\begin{comment}
For example, a codeword designed by the algorithm in \cite{liang2017phase} is shown in Fig. \ref{ps-2d} 
\begin{figure}[h]
    \centering
    \includegraphics[width=2.5in]{1-3d}
    \caption{Beam pattern of a codeword in 2D BPS codebook}
    \label{ps-2d}
\end{figure}
\end{comment}
Another solution for 2D antenna arrays is using beam-steering codewords, where $\boldsymbol{w}=\frac{1}{\sqrt{N}}\boldsymbol{v}(\theta,\phi)$. However, unlike ULA, the shape of covered space is much more complicated and changes with different steering angles. When steering angle is around the center, the covered space is more like a circle. When beams are steered towards the edge direction, the covered space has irregular shape. Thus, choosing the peak directions of beam-steering vectors is particularly challenging in 2D. Choosing the peak directions are out of scope of this work, however, two examples of steering angles for 3 codewords and 4 codewords are listed in Table  \ref{tars}. 
\begin{table}[!h]\label{table1}
\caption{beam-steering codebooks with 3 and 4 codewords for $2\times2$ UPA} 
\centering
\begin{tabular}{|c|c|c|c|}
\hline
  number of codewords&   beam-steering angles $(\theta,\phi )$\\[1ex] 
 \hline
 3 &  [(90.0, 35.3),(120.0, 19.5),(60.0, 19.5)]\\[1ex] 
 \hline
 4&   [(90.0, 0.0),(41.4, 40.9),(90.0, 60.0),(138.6, 40.9)]\\[1ex] 
 \hline
\end{tabular}
\label{tars}
\end{table}
The beam patterns of the beam-steering codebooks, given in Table \ref{tars}, are shown in Figs. \ref{ps3} and \ref{ps4}. However, the extension of the beam-steering codebooks to a larger number of antennas and larger codebook sizes is not straightforward.

The drawbacks of the mentioned codebooks, and more importantly, the lack of a general codebook design for 2D arrays, motivate us to propose a general framework encompassing different array shapes with different numbers of elements and various performance metrics.
\section{The New Codebook Design Framework}\label{framework}
\subsection{Preliminaries}
One of the closely related topics to our problem, i.e., codebook design for analog beamforming, is the quantization problem, particularly, the vector quantization. A vector quantizer of dimension $N$ and size $K$ is a mapping from a vector in N-dimensional Euclidean Distance $\mathbb{R}^N$ into a finite set $\mathbb{C}$ containing $K$ output or reproduction points. We can write it mathematically as $Q: \mathbb{R}^N \rightarrow \mathbb{C}$ where $\mathbb{C}=\{\mathbb{C}[1],\cdots, \mathbb{C}[K]\}$ and $\mathbb{C}[k]\in \mathbb{R}^N$. The set $\mathbb{C}$ is called the codebook, and $\mathbb{C}[k]$, $1\leq k \leq K$, are called the code vectors or codewords. To measure the vector quantizer performance, a distortion measure $d(s, Q(s))$ has to be defined in association with any input vector $s$ and its reproduction vector $Q(s)$, and consequently the performance of a vector
quantizer can be quantified by average distortion as $J=E_s\{d(s,Q(s))\}$ \cite{katsavounidis1994new}. To
permit tractable analysis and easy evaluation, the distortion measure is often chosen to be the squared error, i.e., $|s-Q(c)|^2$. 

The algorithm often used for generating the codebook is referred to as generalized Lloyd algorithm, since it is a vector generalization of the well-known clustering algorithm due to Lloyd \cite{lloyd1982least}. Application of Lloyd's algorithm can be found in the machine learning contexts as well, such as  Adversarial networks \cite{pezeshkpour2019investigating}, GANs \cite{srinivasan2019generating} and completion/pruning methods \cite{pezeshkpour2018embedding,shahhosseini2019partition,pezeshkpour2018compact}. Generalized Lloyd algorithm is an iterative algorithm based on a set of training vectors  which can be described  as follows:
\begin{itemize}
\item \textbf{Initialization:} The initialization step involves choosing the starting codebook of vectors, i.e.,  $\mathbb{C}_0=\{\mathbb{C}_0[1],\cdots, \mathbb{C}_0[K]\}$
\item 	\textbf{Partitioning:} Given a codebook $\mathbb{C}_i$ obtained from the $i$th iteration, find the optimal partitioning of
the space $\mathbb{R}^N$ using the least-distortion condition, i.e.
\begin{flalign}
\Omega_i[k]=\left\{s| \ \ |s-\mathbb{C}_i[k]|^2<|s-\mathbb{C}_i[j]|^2, j\in \{1,\cdots,K\}-\{k\} \right\}
\end{flalign}
\item \textbf{Codeword-Updating:} For a given partitioning $\{\Omega_i\}$, find the so-called centroids of each cell, i.e.
\begin{flalign}
\mathbb{C}_{i+1}[k]=\argmin_{c} E_{s|s\in \Omega_i[k]}\left\{|s-c|^2 \right\},\  k=1,\cdots,K
\end{flalign}
The solution to the above optimization problems can be obtained as $
\mathbb{C}_{i+1}[k]=E[s|s\in \Omega_i[k]],  k=1,\cdots, K$ \cite{yair1992competitive}. For a finite training set, the expected value is replaced by the sample mean of the corresponding cell.
\item iterate until the convergence criteria is satisfied. 
\end{itemize}
Note that many different initialization methods have been proposed, including random coding, pruning, pairwise nearest-neighbor design, product code, and splitting. A thorough survey of these methods can be found in \cite{gersho2012vector}. In addition, to improve the performance of the vector quantization codebook, methods
other than the standard generalized Lloyd algorithm have been examined, e.g., using stochastic relaxation, such as the simulated annealing \cite{zeger1992globally}. Next, our codebook design algorithm for analog beamforming is introduced.
\subsection{Proposed Algorithm}\label{algorithm}
\begin{figure}[h]
    \centering
    \includegraphics[width=3in]{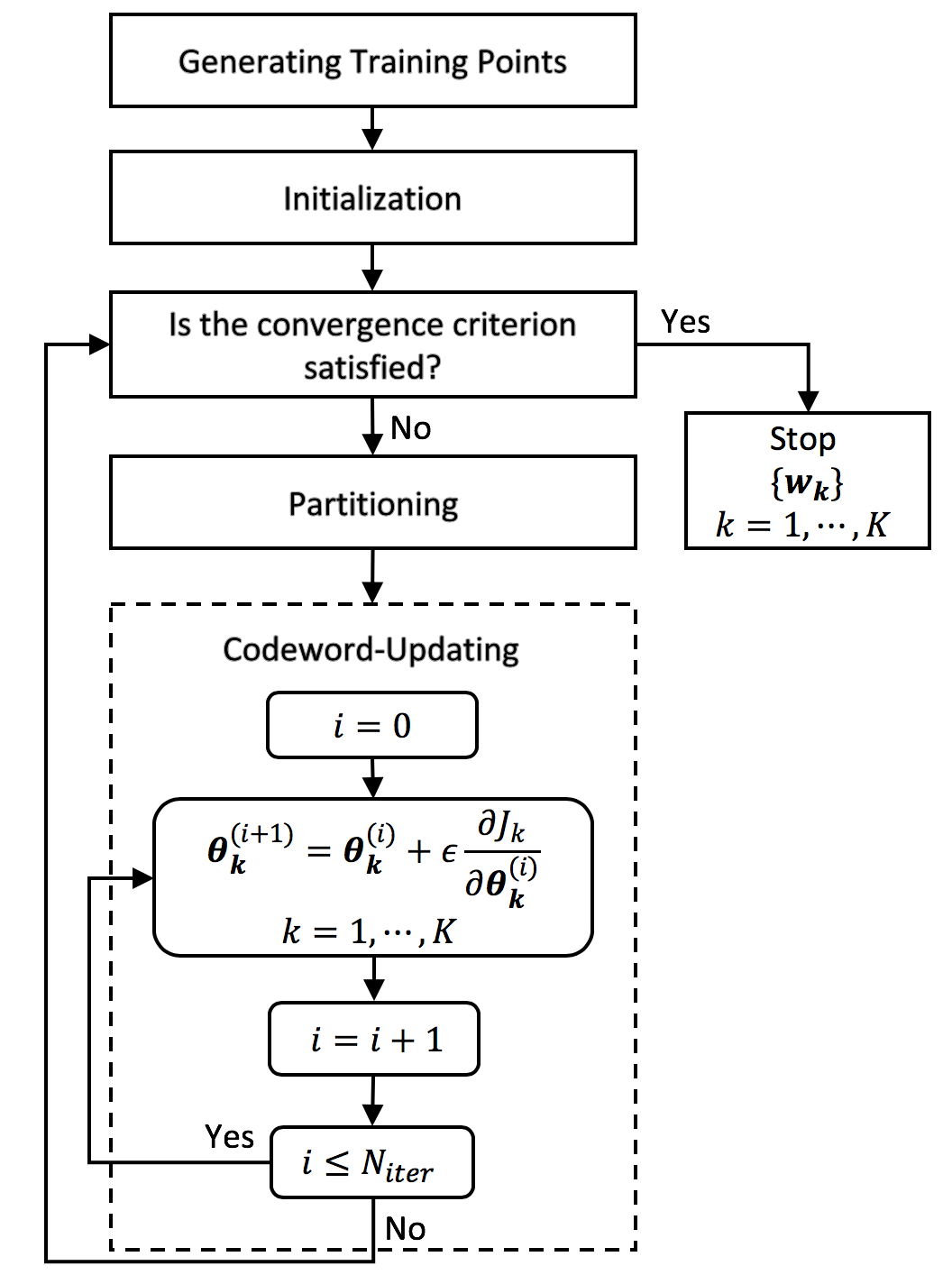}
    \caption{Flowchart of the proposed algorithm}
    \label{flowchart}
\end{figure}

After brief description of the Generalized Lloyd algorithm, the time is ripe to explain our proposed method. We design an algorithm for maximizing a general performance metric defined as follows:
\begin{flalign}
 J=E_{\boldsymbol{h}}\left[f\left(\max_{k=1,\cdots,K}\left|{\mathbb{W}[k]^H\boldsymbol{h}}\right|^2\right)\right]
 \label{form}
\end{flalign}
The algorithm can be described as follows:
\begin{itemize}
\item \textbf{Generating Training Points}: Generate  L  training vectors of $\boldsymbol{h_l}, \ \ l=1,\cdots, L$. 

\item \textbf{Initialization}: Generate initial codewords, $\mathbb{W}_0=\{\mathbb{W}_0[1],\cdots, \mathbb{W}_0[K]\}$

\item \textbf{Partitioning}: Partition all the training vectors into $K$ cells, such that $\Omega_i[k]=\left\{\boldsymbol{h}|\ \ |\mathbb{W}_i[k]^H\boldsymbol{h}|^2> |\mathbb{W}_i[j]^H\boldsymbol{h}|^2 , j\in \{1,\cdots,K\}-\{k\}\right\}$
	
 \item \textbf{Codeword-Updating}: 
 To update the codewords in each iteration, we need to maximize $J[k]=\frac{1}{|\Omega[k]|}\sum_{\boldsymbol{h} \in \Omega[k]}f\left(|\boldsymbol{w}^H\boldsymbol{h}|^2\right)$. In other words:
\begin{flalign}
\mathbb{W}_{i+1}[k]=\argmax_{\boldsymbol{w}} \frac{1}{|\Omega_i[k]|}\sum_{\boldsymbol{h} \in \Omega_i[k]}f\left(|\boldsymbol{w}^H\boldsymbol{h}|^2\right) , \ \ s.t.\ \  \boldsymbol{w} \in W(N)
\end{flalign}
where $f(.)$ can be any desired metric function. Unfortunately, there is no closed-form solution for a general function, thus we use the Gradient decent method to update the codewords in each iteration. Note that this method can work with any differentiable function. The derivative of $J[k]$ with respect to the vector $\boldsymbol{\theta}$, can be denoted as:
\begin{flalign}
\frac{\partial J[k]}{\partial \boldsymbol{\theta}}= \frac{1}{|\Omega[k]|}\sum_{\boldsymbol{h} \in \Omega[k]}\frac{\partial f}{\partial x}\frac{\partial x}{\partial \boldsymbol{w}}\frac{\partial \boldsymbol{w}}{\partial \boldsymbol{\theta}}
\end{flalign}
where $x$ is the input variable of function $f(.)$. After some calculations, $\frac{\partial J[k]}{\partial \boldsymbol{\theta}}$ can be calculated as:
\begin{flalign}
\frac{\partial J[k]}{\partial \boldsymbol{\theta}}= \frac{1}{|\Omega[k]|}\sum_{\boldsymbol{h} \in \Omega[k]} f'\left(|\boldsymbol{w}^H\boldsymbol{h}|^2\right)\boldsymbol{w}^H\boldsymbol{h}\boldsymbol{h}^H\boldsymbol{\Theta}
\end{flalign}
where $\boldsymbol{\Theta}$ is a diagonal matrix whose $n$th diagonal element is equal to $je^{j\boldsymbol{\theta}(n)}$. Then, the phase shifts of the $k$th codeword can be updated as $\boldsymbol{\theta_k}^{(i+1)}=\boldsymbol{\theta_k}^{(i)}+\epsilon \left.\frac{\partial J[k]}{\partial \boldsymbol{\theta_k}}\right|_{@\boldsymbol{\theta_k}^{(i)}}$ where $\epsilon$ is the step size and should be tuned carefully along with the number of iterations.  
\item Iterate until convergence criteria is satisfied.
\end{itemize}
\textbf{Remarks:}
The training points can be empirical points obtained by measurements or drawn from some distributions based on the characteristics of the system. If no prior information is available, uniform spatial coverage, i.e., single-ray with uniform angle of arrival (AOA), can be considered. The proposed algorithm is a decent algorithm, i.e., at each iteration in the algorithm a positive improvement in performance is achieved over the previous iteration, which converges to a local optimal. Powerful initialization techniques and optimization methods that introduce elements of randomness into the algorithm can be applied to improve the resulting performance \cite{zeger1992globally}; however, in this work, to disclose the main concepts, we simply use random initialization and standard gradient decent method as described above. Further improvements in the proposed algorithm can be considered for future work. The algorithm iterates until some convergence criteria, e.g., the difference between two consecutive objective metrics, is satisfied. The flowchart of the algorithm is shown in Fig. \ref{flowchart}.

The proposed algorithm scheme has several attractive
properties. First, it can be applied to arbitrary codebook sizes. Second, it is indifferent to the training channel vectors and it can be applied to different array shapes including ULA, UPA and UCA. Finally, it can be employed to optimize various performance metrics, e.g., average beamforming gain ($f(x)=x$), average data rate $(f(x)=log(1+x))$ and outage ($f(x)=sigmoid(x-\gamma)$). 
\vspace{-10pt}
\subsection{Outage/Coverage Analysis}
\begin{figure}[!h]
\centering
\subfloat[The LB codebook optimizing $J_{avg}$: $J_{out}(8)=19\%$]{ 
\includegraphics[width=2in]{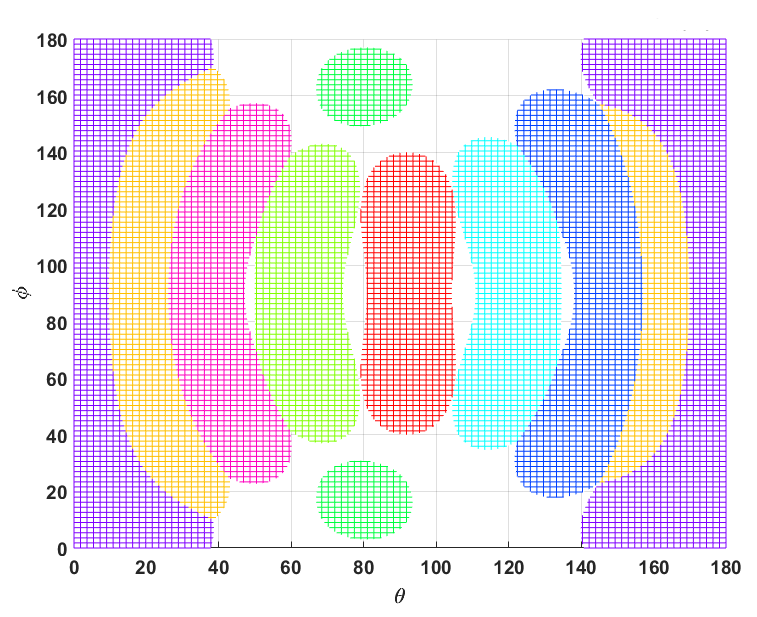}
\label{3db-1}}
\hfil
\subfloat[The LB codebook optimizing $J_{out}(8)$: $J_{out}(8)=15\%$]{
\includegraphics[width=2in]{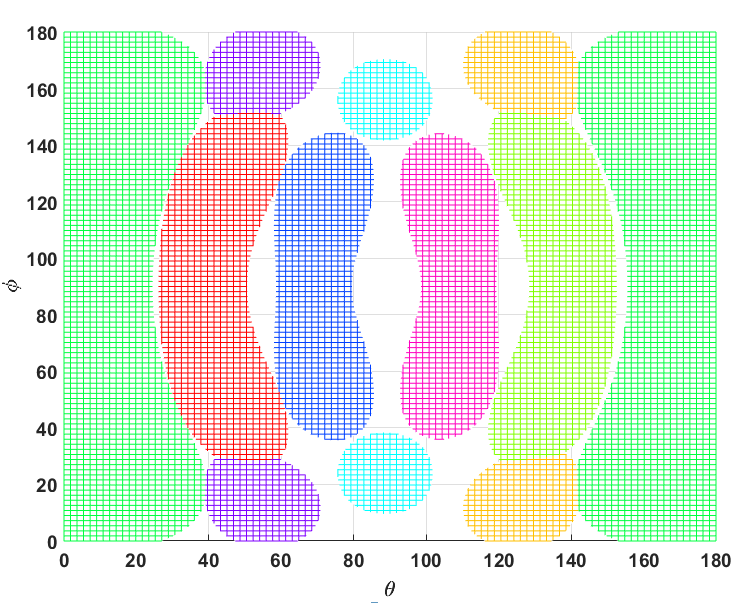}
\label{3db-2}}
\caption{3dB-coverage}
\label{3db}
\end{figure}

\begin{figure}[!h]
\centering
\subfloat[The LB codebook optimizing $J_{avg}$: $J_{out}(4)=7\%$]{ 
\includegraphics[width=2in]{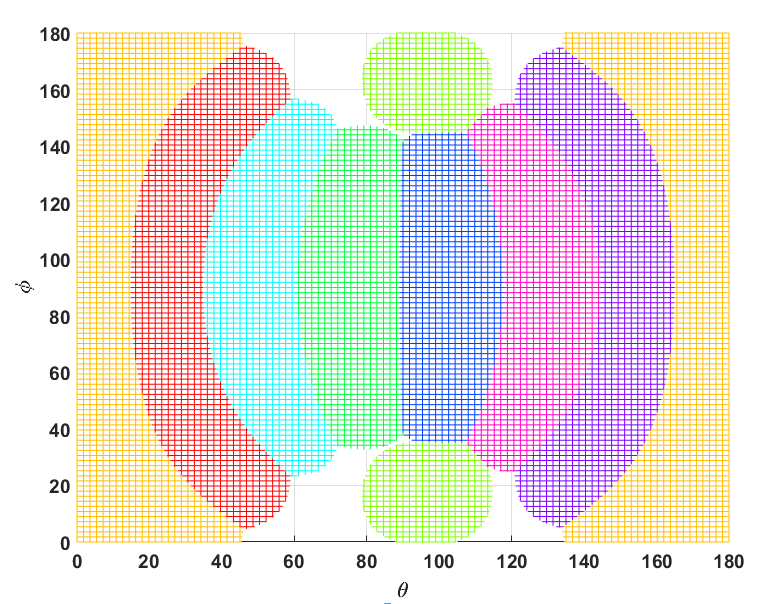}
\label{6db-1}}
\hfil
\subfloat[The LB codebook optimizing $J_{out}(4)$: $J_{out}(4)=0.1\%$]{
\includegraphics[width=2in]{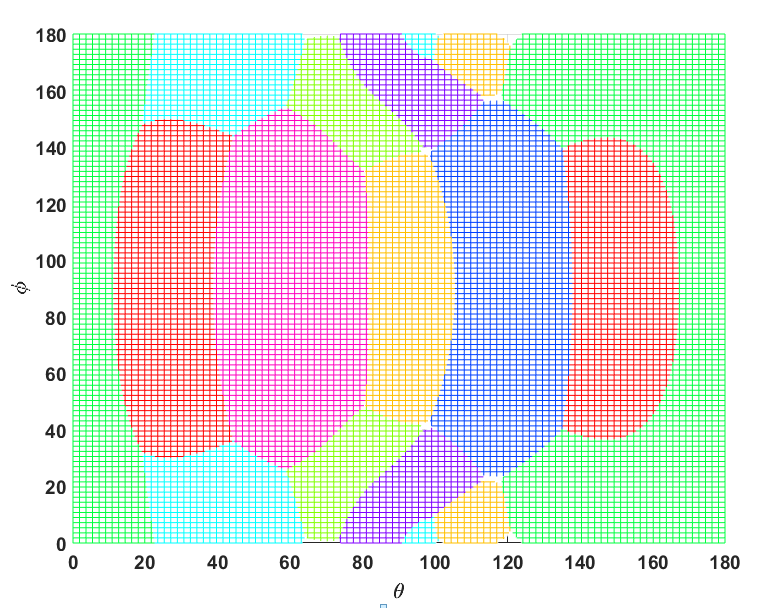}
\label{6db-2}}
\caption{6dB-coverage }
\label{6db}
\end{figure}

Here, the effect of the optimization metric in the effective spatial response is shown. In Figs. \ref{3db-1} and \ref{3db-2}, the effective spatial response of two codebooks with 8 codewords are shown which are designed to optimize $J_{avg}$ and $J_{cov}(8)$, respectively, for a $4\times 4$ UPA antenna array. The 3dB-outage of the two codebooks optimizing $J_{avg}$ and $J_{out}(8)$ are $19\%$ and $15\%$, respectively. In Figs. \ref{6db-1} and \ref{6db-2}, the 6dB-coverage of the codebooks optimizing $J_{avg}$ and $J_{out}(4)$ are shown., respectively. The 6dB-outage of the two codebooks are $7\%$ and $0.1\%$, verifying the effectiveness of the proposed algorithm for minimizing the outage. More simulation results are shown later, in Section \ref{simulation}.
\subsection{Quantized Phase Shifters}
So far, we have assumed ideal phase shifters with no constraint on the precision of introduced phase shifts. However, in practice, phase shifters might be limited to few quantized values. Assume that the value of each phase shifter is specified by $B$ bits. Then, the codebook vectors can only be chosen from $2^{NB}$ feasible quantized vectors. In other words, the feasible set, i.e., $W(N)$, with an infinite number of entries, reduces to a quantized set with $2^{NB}$ entries. The most obvious solution for designing the desired codebook with $K$ codewords is an exhaustive search of all $\binom{2^{NB}}{K}$ combinations of feasible vectors to select the best combination that optimizes our desired metric. However, the complexity of the exhaustive method hinders the practicality of this method. Therefore, to design the codebook, we combine the projection approximation with the proposed algorithms in the previous section. At the end of each iteration, we project the found codewords to the closest vectors in the feasible set of quantized vectors. Furthermore, to avoid divergence, we only update the codeword when the new feasible option is an improvement over the previous feasible option.
\subsubsection{Quantized Codebook for Optimizing $J_{avg}$}
In Fig. \ref{ideal_quant_avg}, the effect of quantization in the average beamforming gain is shown. A $2\times2$ UPA array with 4 codewords and a $4\times4$ UPA array with 8 codewords is considered. To achieve the performance of ideal codebook with no quantization, 5 bits per antenna is required. The performance of 1-bit phase shifters depends on the number of antennas in the array. For example, in a $4\times4$ UPA array $12\%$ and in a $2\times2$ UPA array $37\%$ loss occurs, thus, large number of antennas compensates for the loss of quantization and very low resolution phase shifters can enjoy negligible loss when the array size is large. Further analysis are left for future work.
\subsubsection{Quantized Codebook for Optimizing the Outage Metric}
\begin{figure}[!h]
\centering
\subfloat[Average performance metric]{ 
\includegraphics[width=3in]{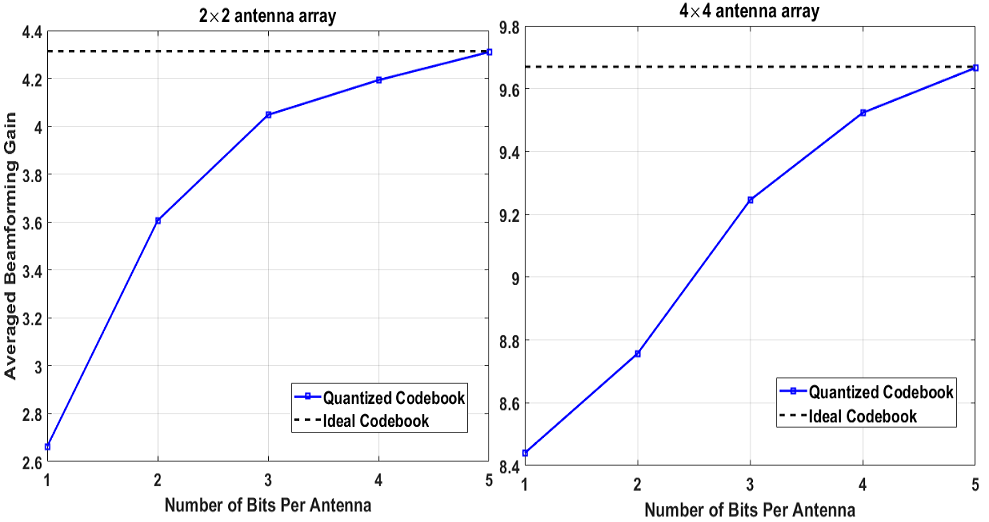}
\label{ideal_quant_avg}}
\hfil
\subfloat[Outage performance metric]{
\includegraphics[width=3in]{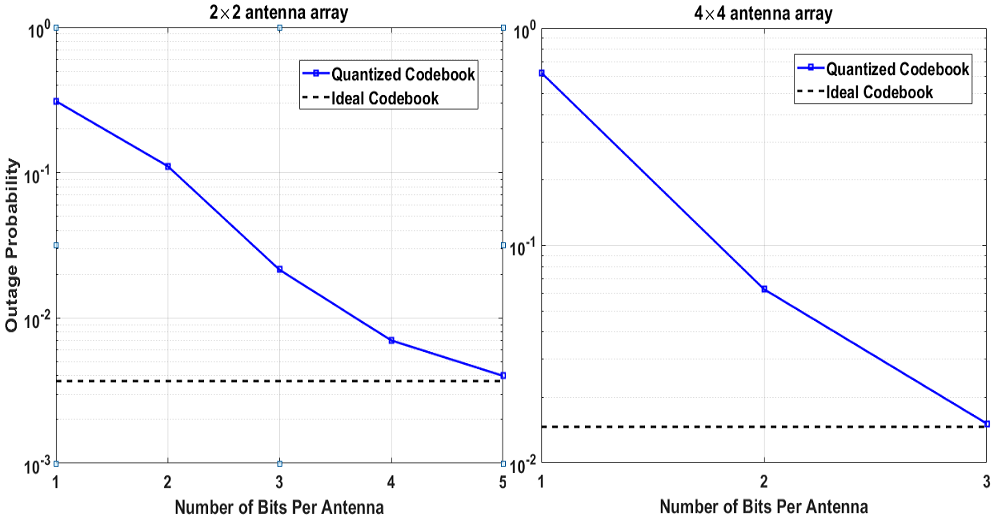}
\label{ideal_quant_outage}}
\caption{Comparison of quantized codebook with ideal codebook }
\label{ideal_quant_comp}
\end{figure}

In Fig. \ref{ideal_quant_outage}, the effect of quantization in the outage probability is shown. As you can see, to reach the ideal outage performance, 3 bits and 5 bits per antenna are required when the array size is $4\times4$ and $2\times2$, respectively. However, 1-bit phase shifter does not provide good outage performance, particularly, when the size of array is large. 
\section{Performance Evaluation}\label{simulation}
Here, we provide extensive simulations to evaluate the performance of our proposed LB codebooks in different scenarios considering ULA/UPA, LOS/NLOS, and ideal/quantized phase shifters, as well as different performance metrics. 
\subsection{Comparison of Averaged Beamforming Gain for ULA}\label{mehrie}
\begin{figure}[!h]
\centering
\subfloat[LOS]{ 
\includegraphics[width=2in]{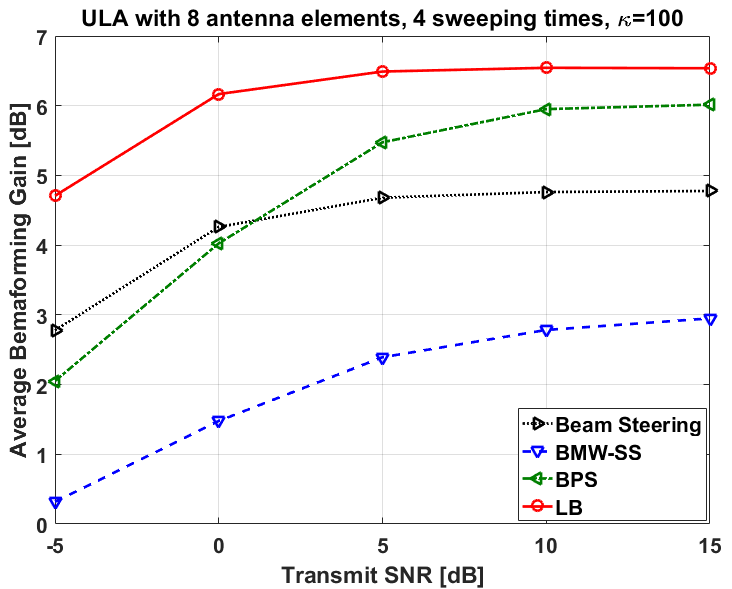}
}
\hfil
\subfloat[NLOS]{
\includegraphics[width=2in]{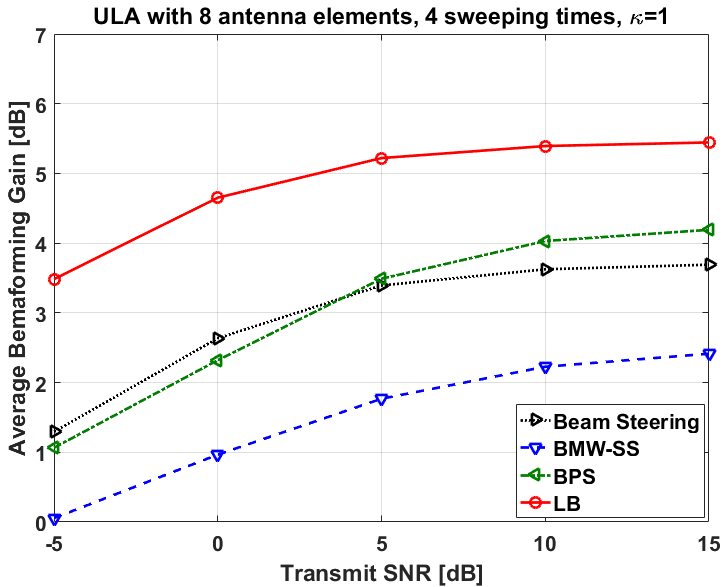}
}
\caption{Performance comparison of ULA with 8 antenna elements and 4 sweeping times}
\label{8_4}
\end{figure}

\begin{figure}[!h]
\centering
\subfloat[LOS]{ 
\includegraphics[width=2in]{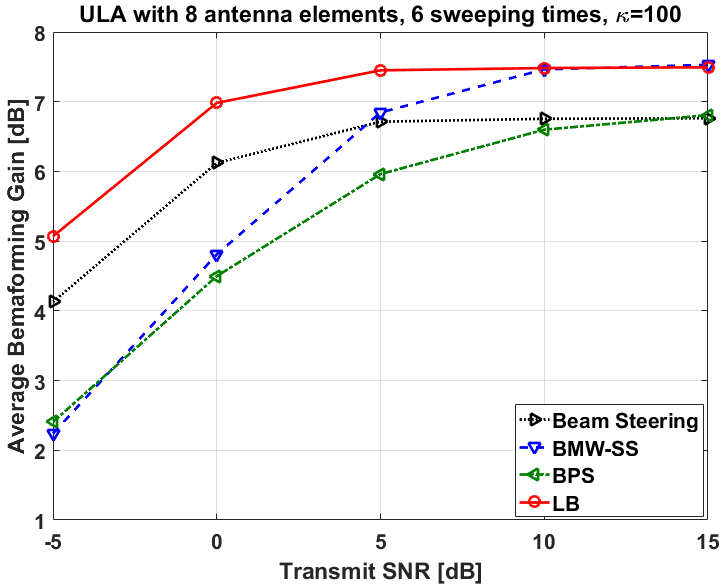}
}
\hfil
\subfloat[NLOS]{
\includegraphics[width=2in]{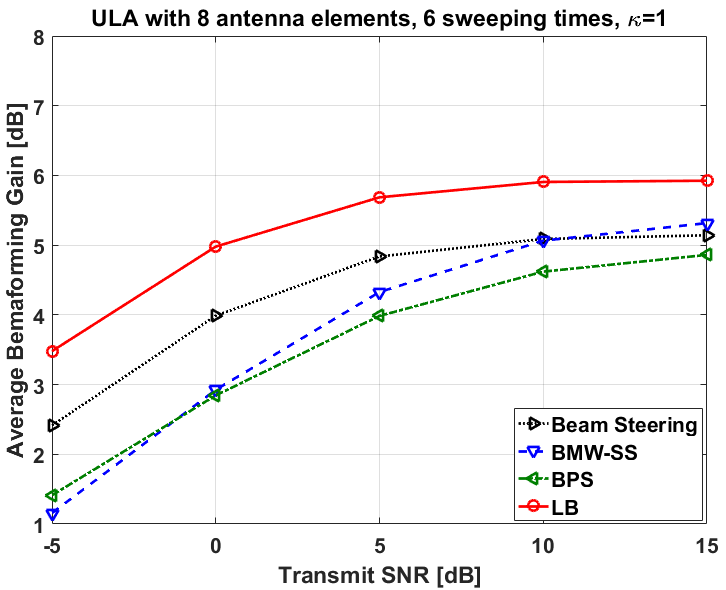}
}
\caption{Performance comparison of ULA with 8 antenna elements and 6 sweeping times}
\label{8_6}
\end{figure}

\begin{figure}[!h]
\centering
\subfloat[LOS]{ 
\includegraphics[width=2in]{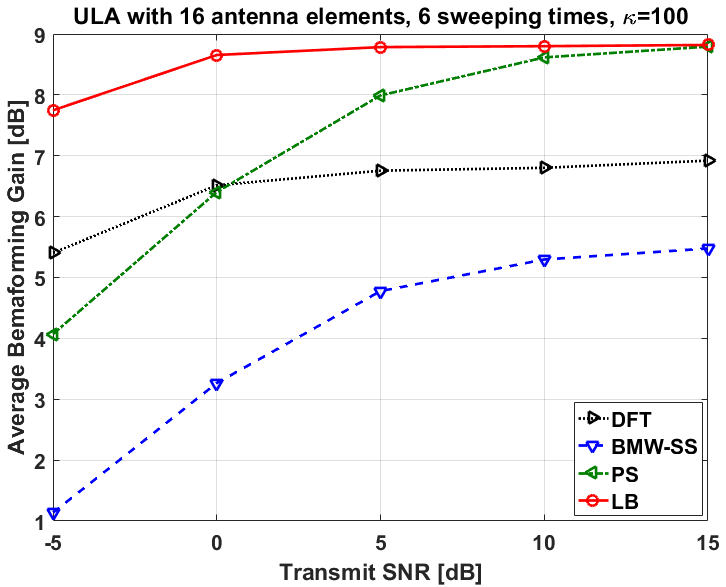}
}
\hfil
\subfloat[NLOS]{
\includegraphics[width=2in]{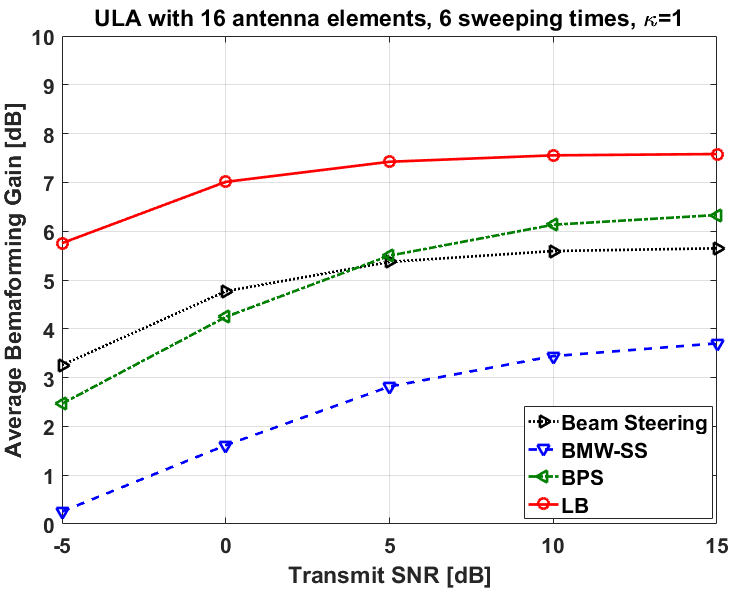}
}
\caption{Performance comparison of ULA with 16 antenna elements and 6 sweeping times}
\label{16_6}
\end{figure}

\begin{figure}[!h]
\centering
\subfloat[LOS]{ 
\includegraphics[width=2in]{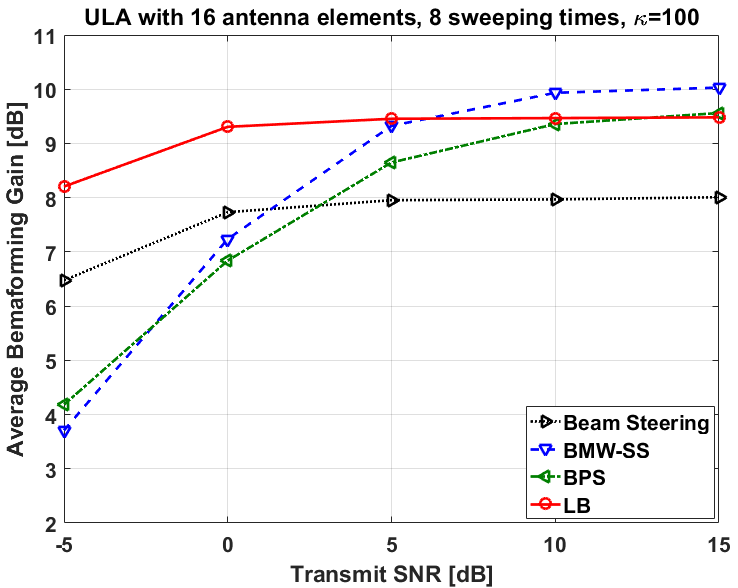}
}
\hfil
\subfloat[NLOS]{
\includegraphics[width=2in]{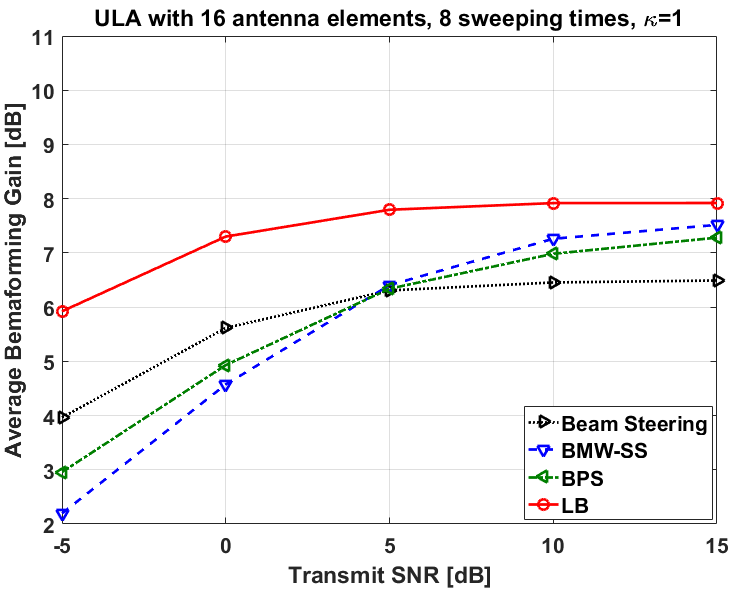}
}
\caption{Performance comparison of ULA with 16 antenna elements and 8 sweeping times}
\label{16_8}
\end{figure}

In this section, the average beamforming gain of different codebooks is compared for ULAs with 8 and 16 antennas. In the simulation set-up, $I=5$ NLOS paths are considered. For LOS and NLOS scenarios, the Ricean factors are assumed to be $100~(20 dB)$ and $1~(0 dB)$, respectively. The average is taken over $10^5$ different realizations (in each realization, $\theta^i_R$ is chosen uniformly from $[0,180]$), and the channel gains and noise samples are drawn from $CN(0,1)$. The beam-steering codebook is created by considering an equi-spaced direction. Assuming $K$ sweeping times, the BMW-SS and BPS codebooks follow a hierarchical design with $K/2$ levels, with two codewords evaluated at each level. For a fair comparison, (4,6) and (6,8) sweeping times are considered for ULAs with 8 and 16 antennas, respectively. The LB codebooks are designed to optimize $J_{avg}$, where the training channel vectors are single-rays whose AOA is uniformly distributed in $[0, 180]$. In Fig. \ref{8_4}, 4 sweeping times are considered; thus, the BMW-SS and BPS codebooks each consist of 2 levels. The LB codebook provides the best average beamforming gain, approximately $4$ dB, $2$dB, and $1$dB more than the BMW-SS, beam-steering, and BPS codebooks, respectively. With a dominant LOS path, the hierarchical designs, particularly the BPS codebook, get closer to the LB codebook in terms of performance. In the low SNR regime, the hierarchical designs suffer from error propagation in the search step for choosing the best codeword, and the performance is even worse than that of the beam-steering codebook.
In Fig. \ref{8_6}, 6 sweeping times are considered; thus, the hierarchical codebooks, i.e., BPS and BMW-SS, contains 3 levels, and the last level of the BMW-SS codebook is the DFT codebook. In the low SNR regime, the LB codebook provides approximately 1 and 2.5 dB gains compared with the beam-steering and BPS/BMW-SS codebooks, respectively, in both, the LOS and NLOS scenarios. In the high SNR regime and LOS scenario, the hierarchical codebook performance is very close to that of the LB codebook, particularly the BMW-SS codebook, whose last level coincides with the DFT codebook. However, in the NLOS scenario, the LB codebook provides better performance.
In Fig. \ref{16_6}, a ULA with 16 antenna elements and 6 sweeping times is considered. The LB codebook provides better performance, especially under the Low SNR regime and NLOS scenario. The BPS codebook can achieve the performance of the LB codebook under the high SNR regime and LOS scenario; however, at low SNR it performs even worse than the beam-steering codebook. The BMW-SS codebook has poor performance because of antenna deactivation. 
In Fig. \ref{16_8}, the BMW-SS codebook slightly outperforms the LB codebook in high SNR and LOS scenario. However, in other scenarios, the LB codebook provides significant gain, particularly in low SNR regime which it provides around 4dB (100\%) gain. 
In summary, the LB codebook provides significant beamforming gain in ULAs, particularly in NLOS scenario and also low SNR regime. Besides the effectiveness of our proposed method for ULAs, it can be easily adapted to other array shapes like UPAs which is evaluated next.  
\subsection{Comparison of Average Beamforming Gain for UPA}
In this section, the average beamforming gain of different codebooks are compared for UPA with $2\times2$ and $4\times4$ number of antennas which are commonly used in 3GPP specifications. Similarly, the average is taken over $10^5$ different realizations of channel, noise vectors, and also $\theta^i_R$ and $\phi^i_R$ which are chosen uniformly from $[0,180]$ and $[-90,90]$, respectively. The Kronecker extension of  BMW-SS and Beam-Sweeping codebooks, 2D BPS codebooks and proposed LB codebooks are compared. hierarchical codebooks evaluate 4 codewords in each level, thus $4$ and $8$ sweeping times is considered. The LB codebooks are designed to optimize $J_{avg}$ where the training channel vectors are single-rays whose AOAs are uniformly distributed in $\theta\sim[0, 180]$ and $\phi\sim[-90,90]$.

\begin{figure}[!h]
\centering
\subfloat[LOS]{ 
\includegraphics[width=2in]{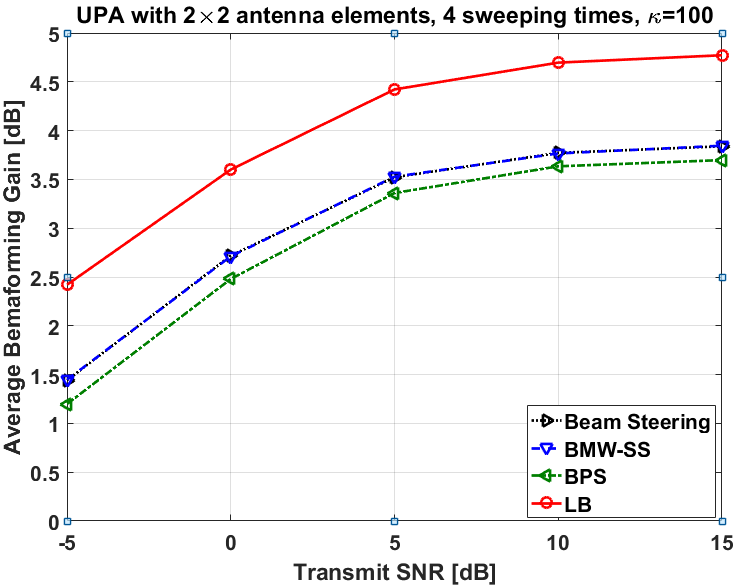}
}
\hfil
\subfloat[NLOS]{
\includegraphics[width=2in]{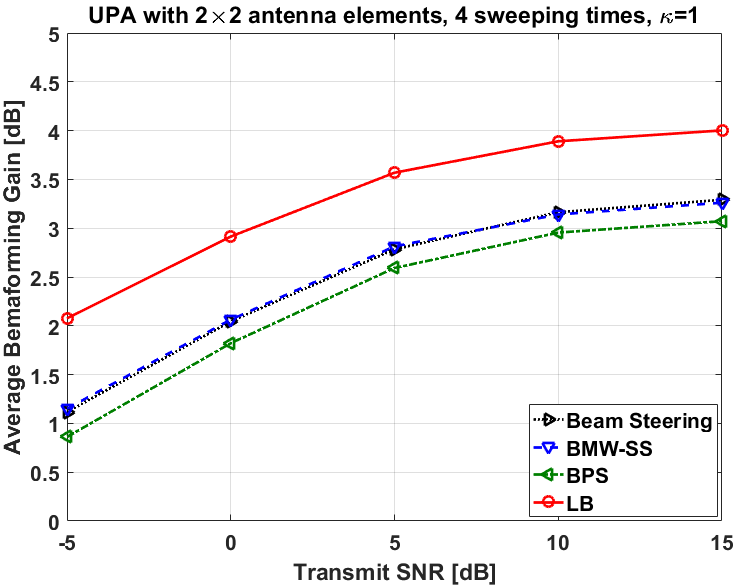}
}
\caption{Performance comparison of UPA with $2\times2$ antenna elements and 4 sweeping times}
\label{2by2_4}
\end{figure}

\begin{figure}[!h]
\centering
\subfloat[LOS]{ 
\includegraphics[width=2in]{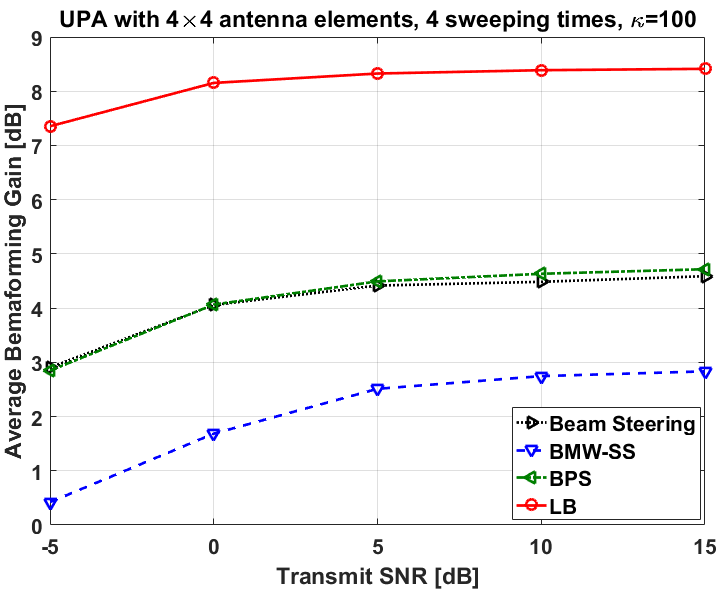}
}
\hfil
\subfloat[NLOS]{
\includegraphics[width=2in]{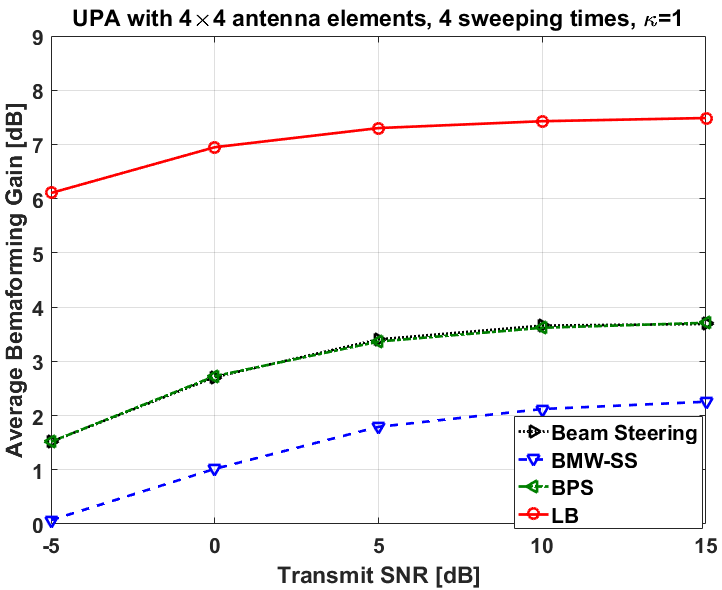}
}
\caption{Performance comparison of UPA with $4\times 4$ antenna elements and 4 sweeping times}
\label{4by4_4}
\end{figure}

\begin{figure}[!h]
\centering
\subfloat[LOS]{ 
\includegraphics[width=2in]{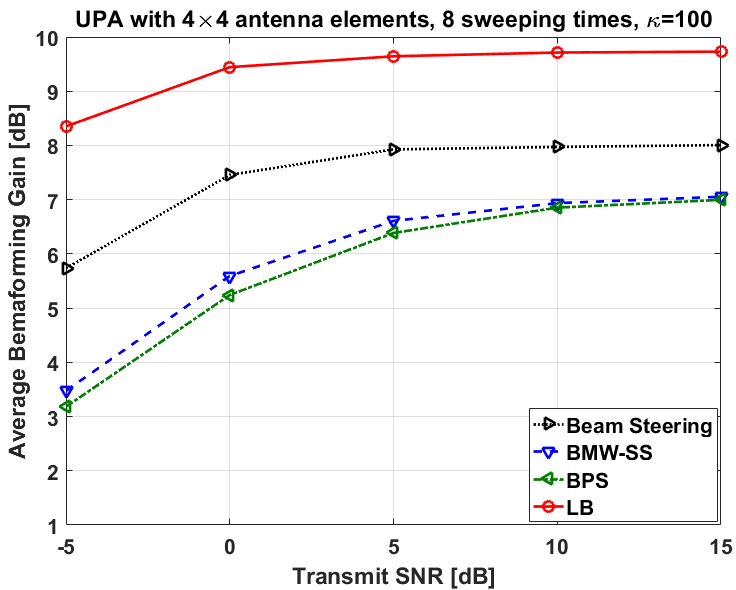}
}
\hfil
\subfloat[NLOS]{
\includegraphics[width=2in]{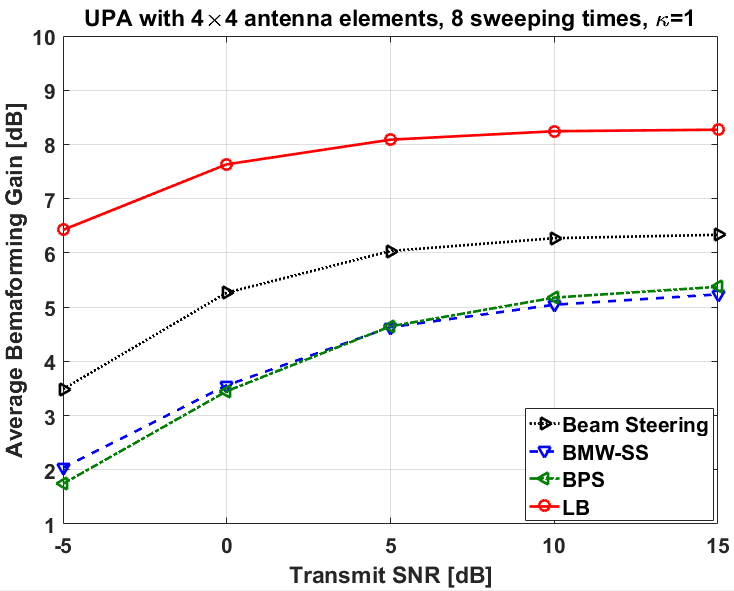}
}
\caption{Performance comparison of UPA with $4\times4$ antenna elements and 8 sweeping times}
\label{4by4_8}
\end{figure}

In Fig. \ref{2by2_4}, $2\times2$ UPA with 4 sweeping times is considered. The LB codebook is providing around 1 dB (100\%) gain compared with other codebooks, in both LOS and NLOS scenarios.  
In Fig. \ref{4by4_4}, $4\times4$ UPA with 4 sweeping times is considered. As it can be seen, noticeable gain is achieved by using the LB codebook. The BMW-SS has the worst performance due to antenna deactivation.
In Fig. \ref{4by4_8}, $4\times4$ UPA with 8 sweeping times is considered. Thus, the BMW-SS, beam-steering and BPS codebooks consists of 2 levels which result in improvement in their performance. However, the LB codebook still outperforms the other codebooks. 
In the next section, the codebooks are compared in terms of their effective spatial responses.
\subsection{Effective Spatial Response}
\begin{figure}[!h]
\centering
\subfloat[2 codewords]{ 
\includegraphics[width=2in]{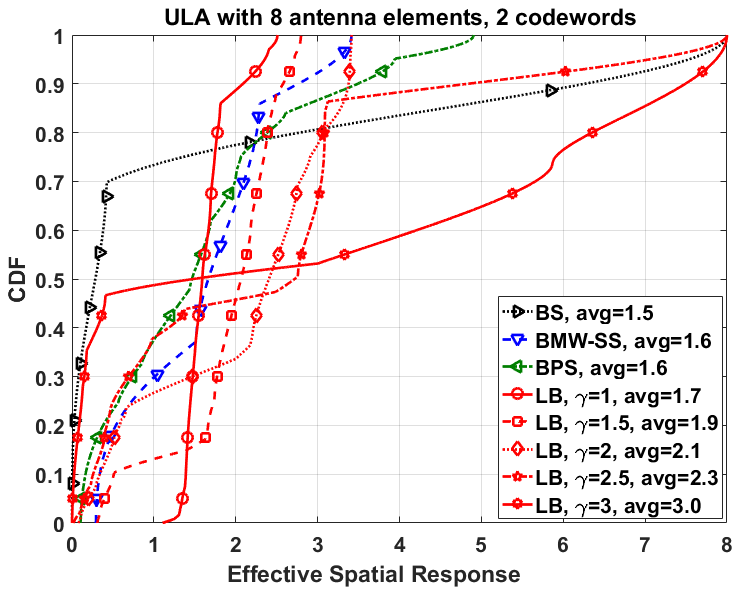}
\label{cdf1}}
\hfil
\subfloat[4 codewords]{
\includegraphics[width=2in]{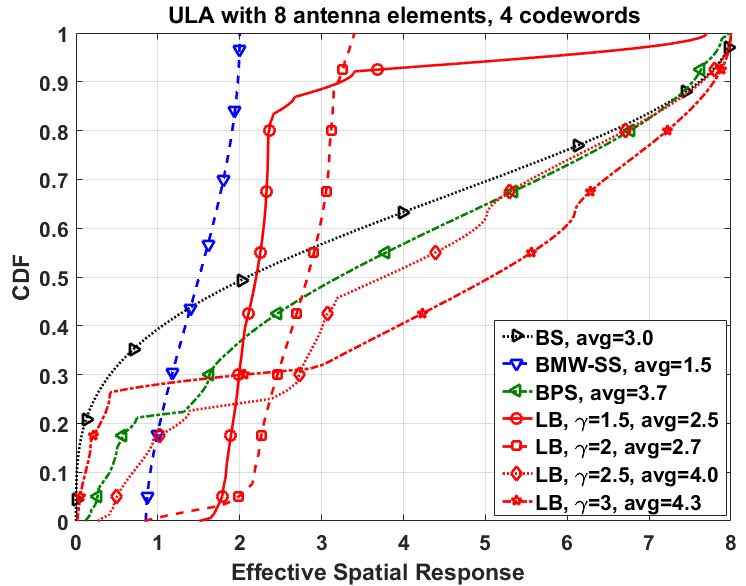}
\label{cdf2}}
\hfil
\subfloat[8 codewords]{
\includegraphics[width=2in]{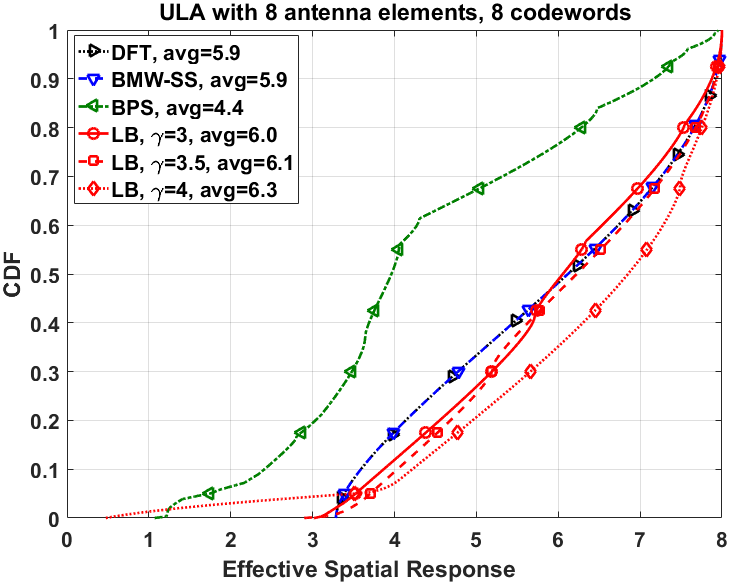}
\label{cdf3}}
\caption{CDF of effective Spatial Response for ULA with 8 antenna elements}
\label{cdf_1d}
\end{figure}

\begin{figure}[!h]
\centering
\subfloat[3 codewords]{ 
\includegraphics[width=2in]{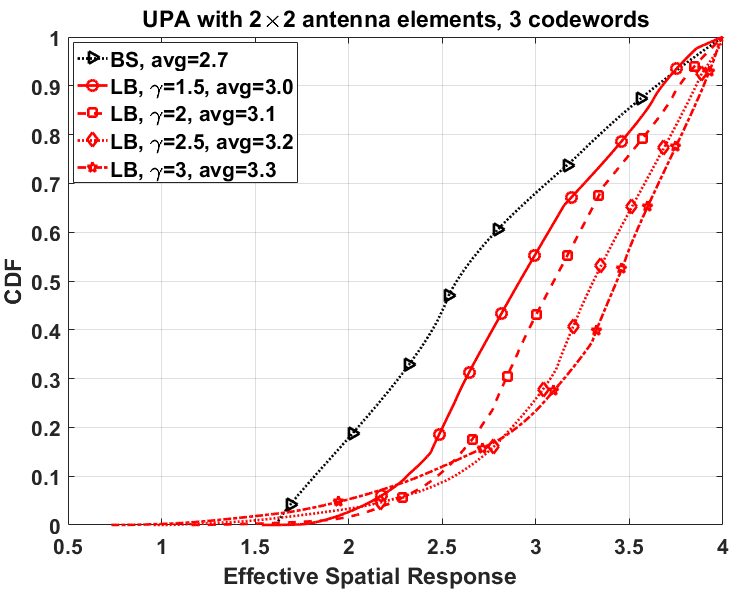}
\label{cdf4}}
\hfil
\subfloat[4 codewords]{
\includegraphics[width=2in]{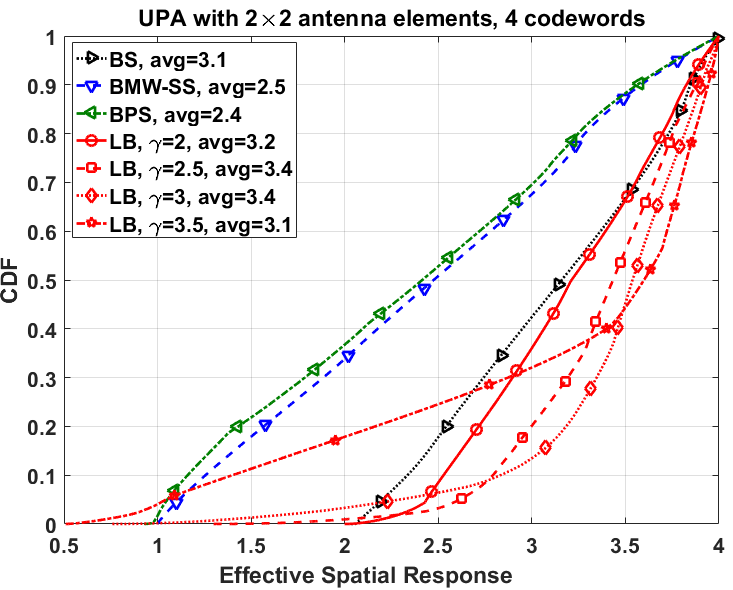}
\label{cdf5}}
\caption{CDF of effective Spatial Response for UPA with $2\times 2$ antenna elements}
\label{cdf_2d}
\end{figure}

\begin{figure}[!h]
\centering
\subfloat[$\gamma=2$]{ 
\includegraphics[width=2in]{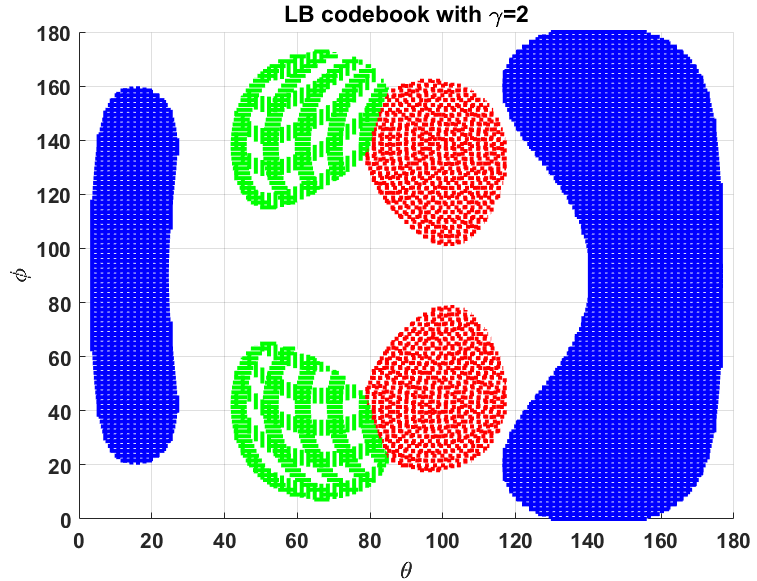}
}
\hfil
\subfloat[$\gamma=2.5$]{
\includegraphics[width=2in]{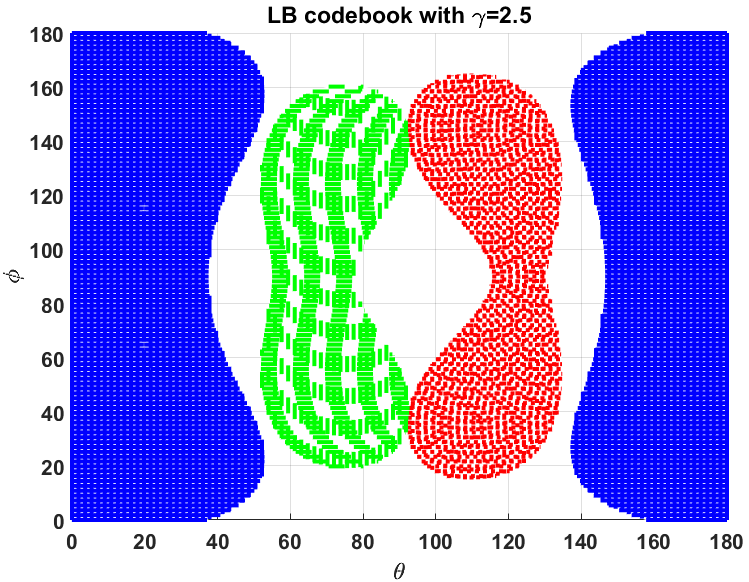}
}
\hfil
\subfloat[$\gamma=3$]{
\includegraphics[width=2in]{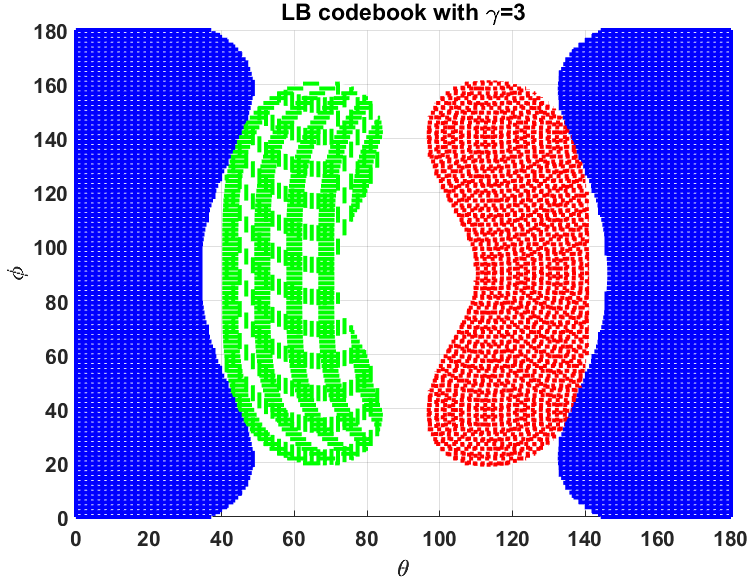}
}
\caption{$3/4$-Coverage of LB codebook with 3 codewords and different outage thresholds}
\label{2by2_3_gamma}
\end{figure}

\begin{figure}[!h]
\centering
\subfloat[$\gamma=2$]{ 
\includegraphics[width=2in]{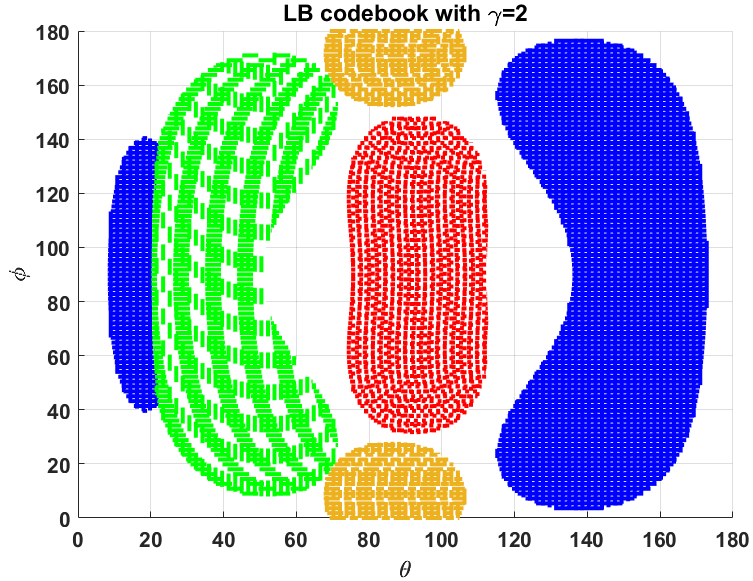}
}
\hfil
\subfloat[$\gamma=2.5$]{
\includegraphics[width=2in]{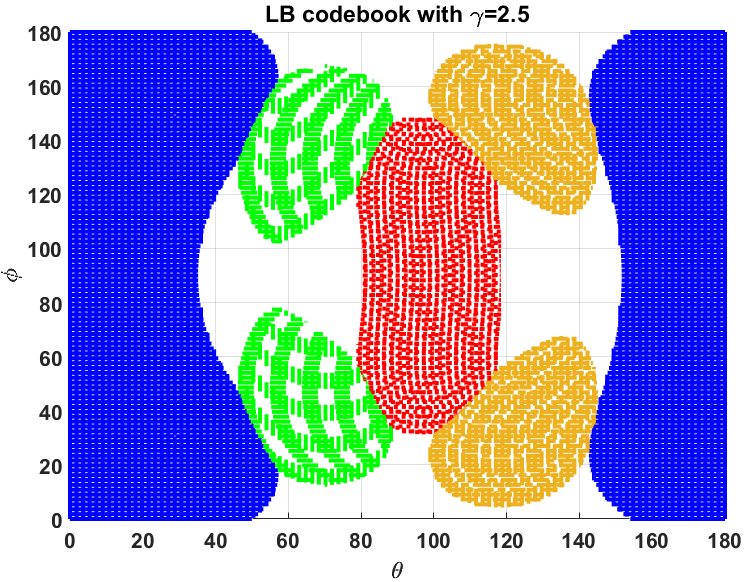}
}
\hfil
\subfloat[$\gamma=3$]{
\includegraphics[width=2in]{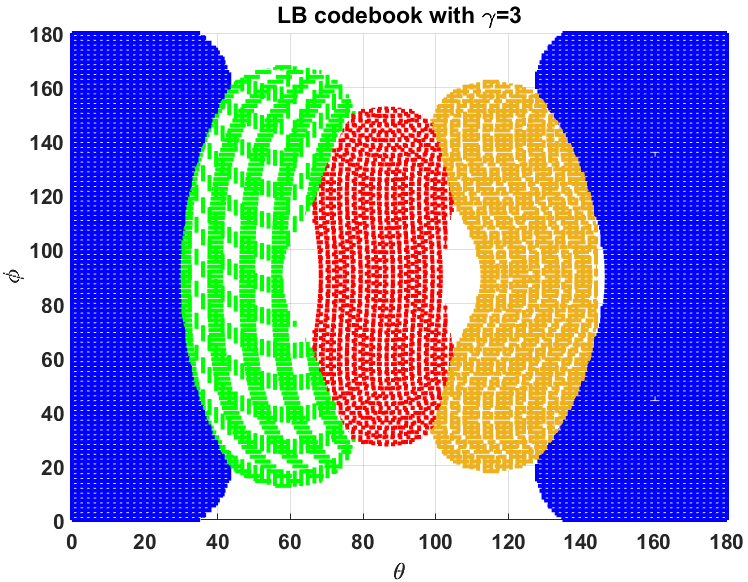}
}
\caption{$3/4$-Coverage of LB codebook with 4 codewords and different outage thresholds}
\label{2by2_4_gamma}
\end{figure}

As explained before, the effective spatial response of a codebook, i.e., $B(\mathbb{W},\theta)=\max_{k=1,\cdots,K}\left|{\mathbb{W}[k]^H\boldsymbol{v}(\theta)}\right|^2$ for ULA and $B(\mathbb{W}, \theta, \phi)=\max_{k=1,\cdots,K}\left|{\mathbb{W}[k]^H\boldsymbol{v}(\theta,\phi)}\right|^2$ for UPA, can be considered as a random variable depending on the codebook $\mathbb{W}$. We compare the CDF of spatial response for different codebooks, assuming $\theta$ and $\phi$ to be uniformly distributed over $[0,180]$. We use CDF to capture all the properties of effective spatial response, as the average or max/min metrics might be unable to show the distinctions between the proposed codebooks. In Fig. \ref{cdf_1d}, ULA with 8 antenna elements with different number of codewords are considered. The LB codebooks in this section are designed to optimize $J_{out}(\gamma)$ for different values of $\gamma$. In Fig.\ref{cdf1}, 2 codewords are considered and all three codebooks of beam-steering, BMW-SS and BPS provide $\overline{B}(\mathbb{W})\approx 1.5$, however, their reliability in performance is different. For example, for both BMW-SS and BPS we have $Pr(B(\mathbb{W})<1)\approx 0.3$, however, for the beam-steering codebook we have $Pr(B(\mathbb{W})<1)\approx 0.73$. Although BMW-SS and BPS codebooks provide more reliability they cannot obtain the maximum gain which can be possibly achieved by the array. Interestingly, the proposed framework provides great flexibility and enables us to design different codebooks satisfying different system requirements.   
The codebook designed to optimize $J_{out}(\gamma)$ provides the least outage at the given threshold, $\gamma$, verifying the effectiveness of our proposed method. For example, in Fig. \ref{cdf1}, where 2 codewords are considered, $J_{out}(1)=0$, $J_{out}(1.5)=15\%$, $J_{out}(2)=33\%$, $J_{out}(2.5)=47\%$ and $J_{out}(3)=53\%$ are obtained by the codebooks designed with the corresponding outage threshold.  The codebook designed with $\gamma=3$ provides the highest average spatial response $\overline{B} (\mathbb{W})=3$. On the other hand, the codebook with $\gamma=1$ provides $J_{out}(1)=0$ while LB codebooks with $\gamma=1.5, 2, 2.5,3$, BMW-SS, BPS and beam-steering codebooks provide $J_{out}(1)=12\%$, $J_{out}(1)=27\%$, $J_{out}(1)=38\%$, $J_{out}(1)=49\%$, $J_{out}(1)=30\%$, $J_{out}(1)=38\%$ and $J_{out}(1)=73\%$, respectively. Thus, the outage threshold can be adjusted based on the system requirements. Same analysis is shown in Figs. \ref{cdf2} and \ref{cdf3} for 4 and 8 number of codewords, respectively. The maximum value of $\overline{B}(\mathbb{W})$ for 2, 4 and 8 codewords is equal to 3, 4.3 and 6.3, respectively. In addition, the maximum value of $\gamma$ such that $J_{out}(\gamma)=0$, is equal to $1.2$, $1.7$ and $3.2$ for 2, 4 and 8 number of codewords, respectively. Our framework can be utilized to design the system parameters, e.g., number of required sweeping resources, such that specific performance metrics are satisfied. 
\begin{figure}[!h]
\centering
\subfloat[$\gamma=1$]{ 
\includegraphics[width=2in]{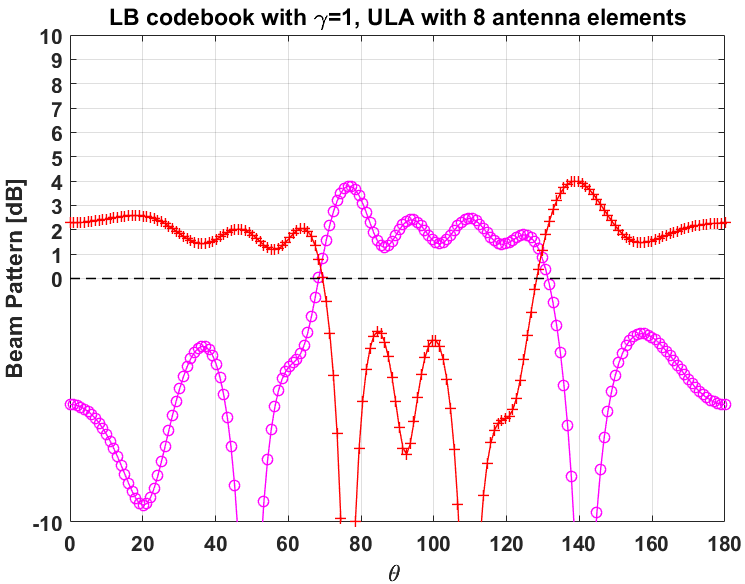}
}
\hfil
\subfloat[$\gamma=2$]{
\includegraphics[width=2in]{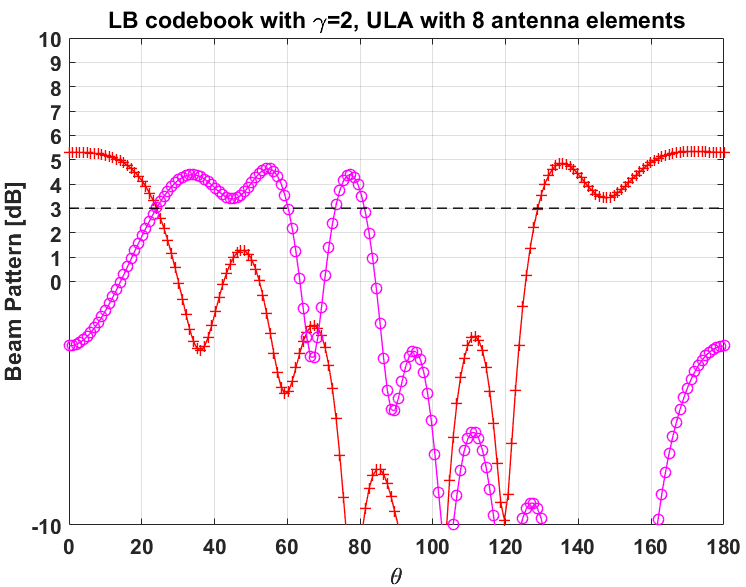}
}
\hfil
\subfloat[$\gamma=3$]{
\includegraphics[width=2in]{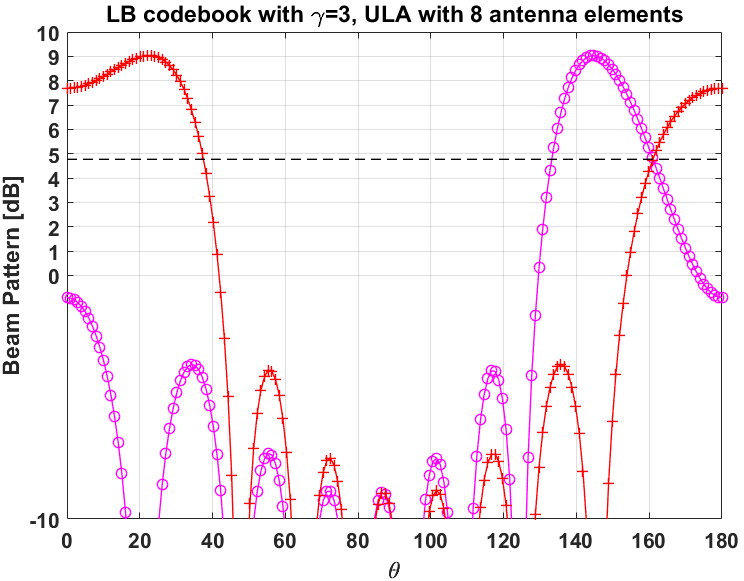}
}
\caption{Beam patterns of LB codebook with 2 codewords and different outage thresholds}
\label{8_2_gamma}
\end{figure}
\begin{figure}[!h]
\centering
\subfloat[$\gamma=1.5$]{ 
\includegraphics[width=2in]{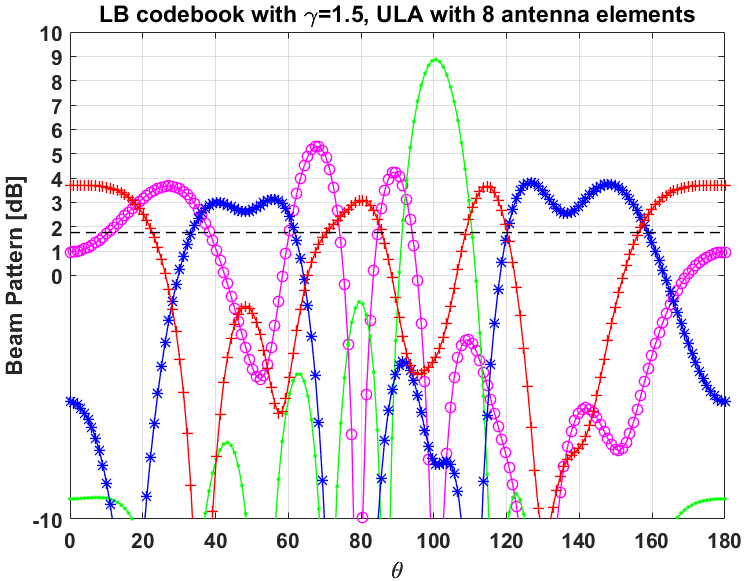}
}
\hfil
\subfloat[$\gamma=2$]{
\includegraphics[width=2in]{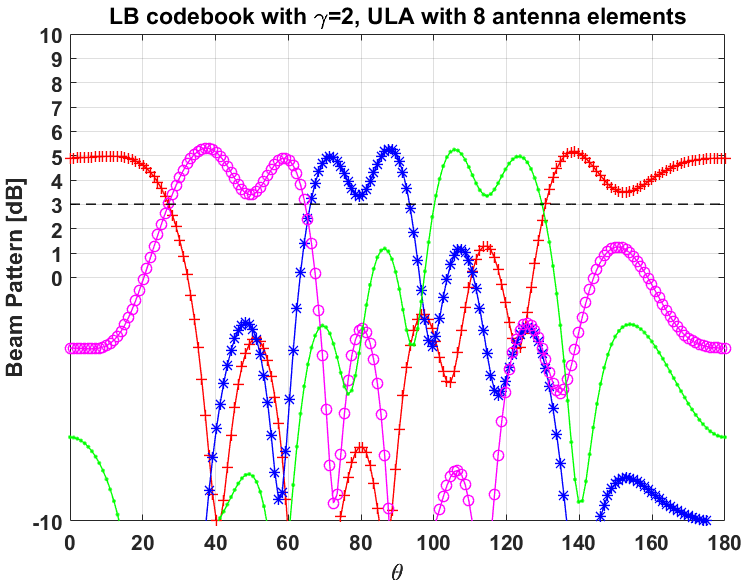}
}
\hfil
\subfloat[$\gamma=3$]{
\includegraphics[width=2in]{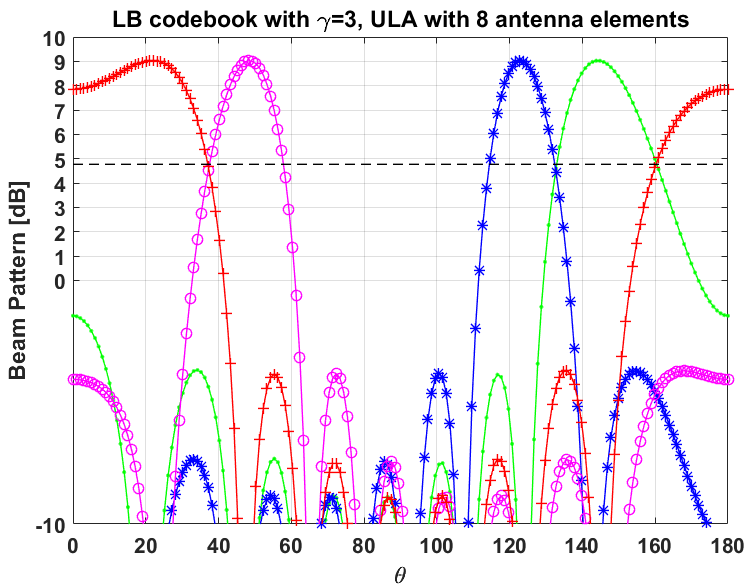}
}
\caption{Beam patterns of LB codebook with 4 codewords and different outage thresholds}
\label{8_4_gamma}
\end{figure}
\begin{figure}[!h]
\centering
\subfloat[$\gamma=3$]{ 
\includegraphics[width=2in]{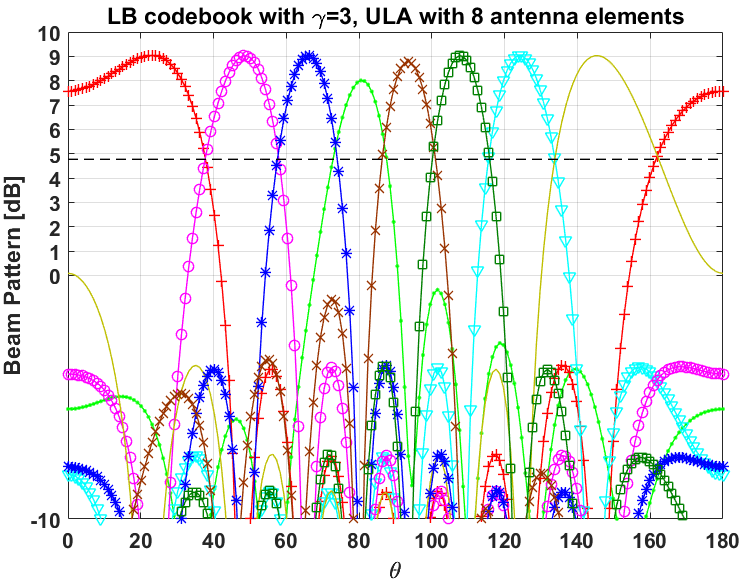}
}
\hfil
\subfloat[$\gamma=3.5$]{
\includegraphics[width=2in]{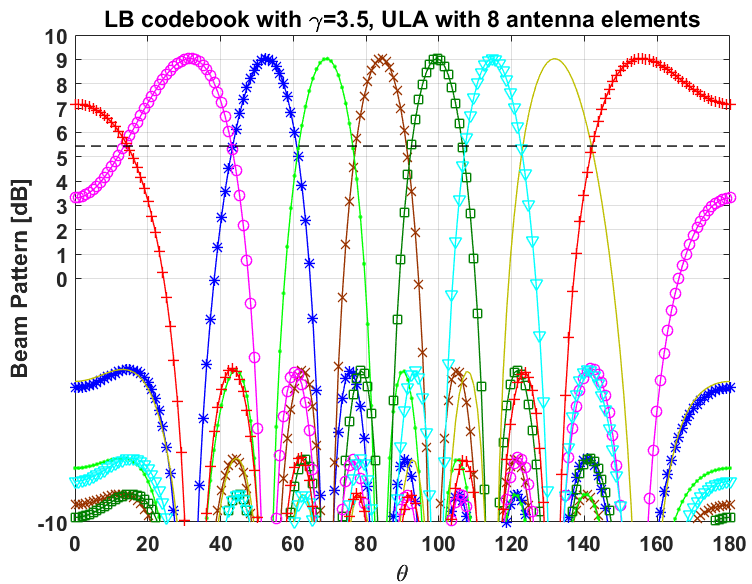}
}
\hfil
\subfloat[$\gamma=4$]{
\includegraphics[width=2in]{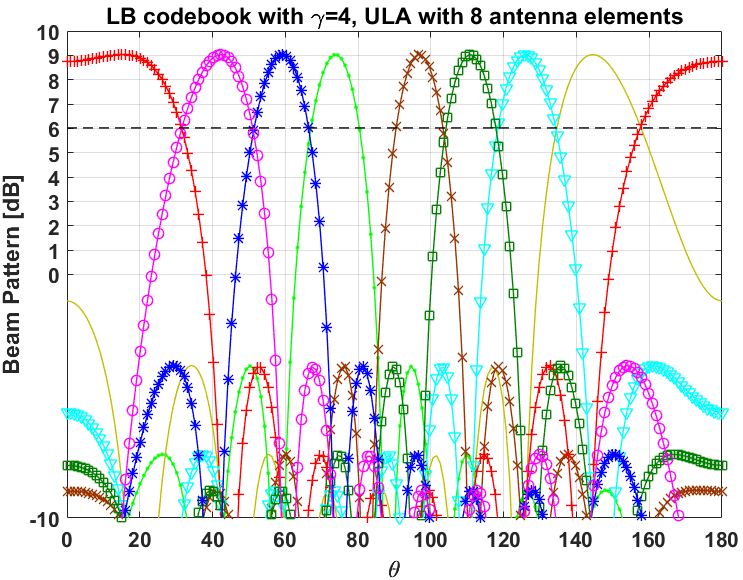}
}
\caption{Beam patterns of LB codebook with 8 codewords and different outage thresholds}
\label{8_8_gamma}
\end{figure}
In addition, the beam patterns of LB codebooks with different values of $\gamma$ are shown in Figs. \ref{8_2_gamma}, \ref{8_4_gamma} and \ref{8_8_gamma}. It can be noted that the beam patterns are flexible and can take irregular shapes unlike the BPS codebook where the beam patterns are dictated to take some pre-defined shapes. In Fig. \ref{cdf_2d}, $2\times 2$ UPA with 3 and 4 number of codewords are considered. In Fig. \ref{cdf4}, the codebook presented in Table $\ref{tars}$ and LB codebooks optimizing $J_{out}(\gamma=1.5,2,2.5,3)$ are considered. In Fig. \ref{cdf5}, the codebook presented in Table $\ref{tars}$, Kronecker extension of  BMW-SS, 2D BPS and LB codebooks optimizing $J_{out}(\gamma=2,2.5,3)$ are considered. The LB codebooks provide better outage performance compared with other codebooks. For example, when 4 codewords are considered, 3dB-outage of LB codebooks is almost zero except the one for $\gamma=3.5$, while that of BPS and BMW-SS codebooks are around $35\%$. 3dB-outage of the proposed beam-steering codebook in Table \ref{tars} also equals zero, however, its outage at higher thresholds is worse than LB codebooks. The maximum value of $\overline{B}(\mathbb{W})$ for 3 and 4 codewords is equal to 3.3 and 3.4, respectively. Besides, the maximum value of $\gamma$ such that $J_{out}(\gamma)=0$ is equal to $1.8$ and $2.2$ for 3 and 4 number of codewords, respectively.
The $3/4$-coverage of LB codebooks with different values of $\gamma$ are shown in Figs. \ref{2by2_3_gamma}, and \ref{2by2_4_gamma}, respectively. Obviously, the codebooks optimizing $J_{out}(3)$ provide the least 3/4-outage of $22\%$ and $12\%$ with 3 codewords and 4 codewords, respectively. 
\subsection{Data Rate}
\begin{figure}[!h]
\centering
\subfloat[LOS]{ 
\includegraphics[width=2in]{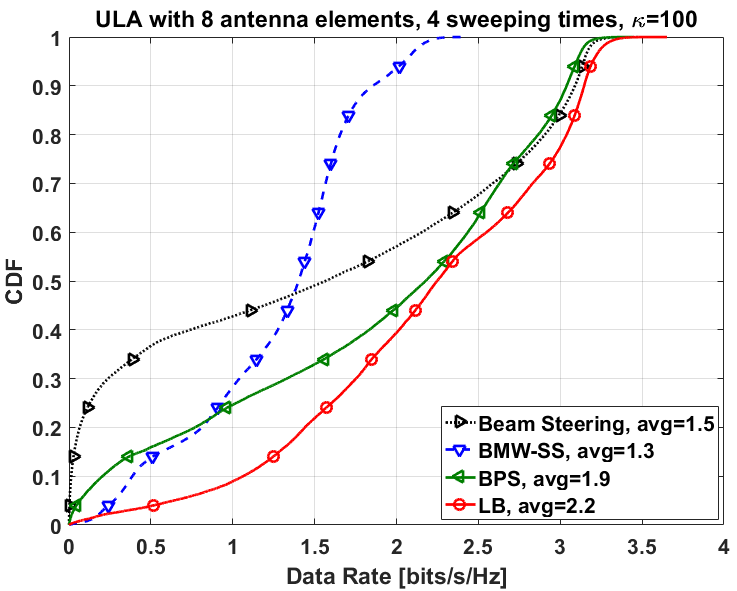}
}
\hfil
\subfloat[N-LOS]{
\includegraphics[width=2in]{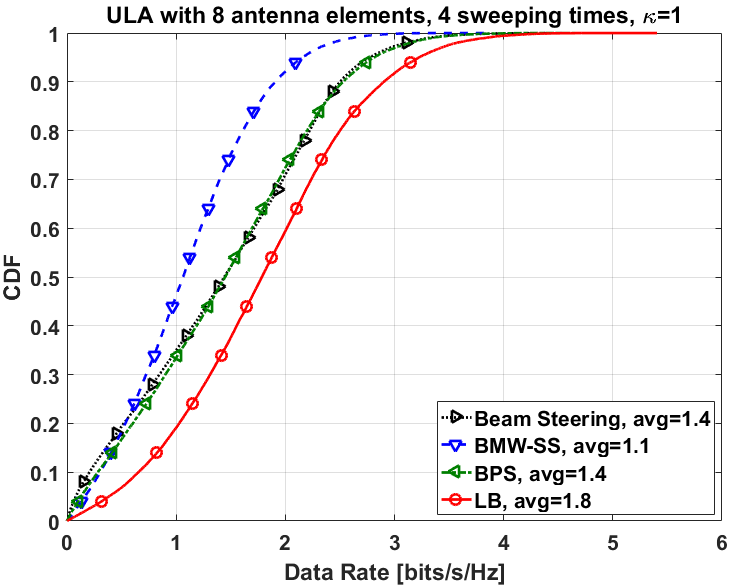}
}
\caption{CDF of Data rate for ULA with 8 antenna elements}
\label{rate_8_4}
\end{figure}

\begin{figure}[!h]
\centering
\subfloat[LOS]{ 
\includegraphics[width=2in]{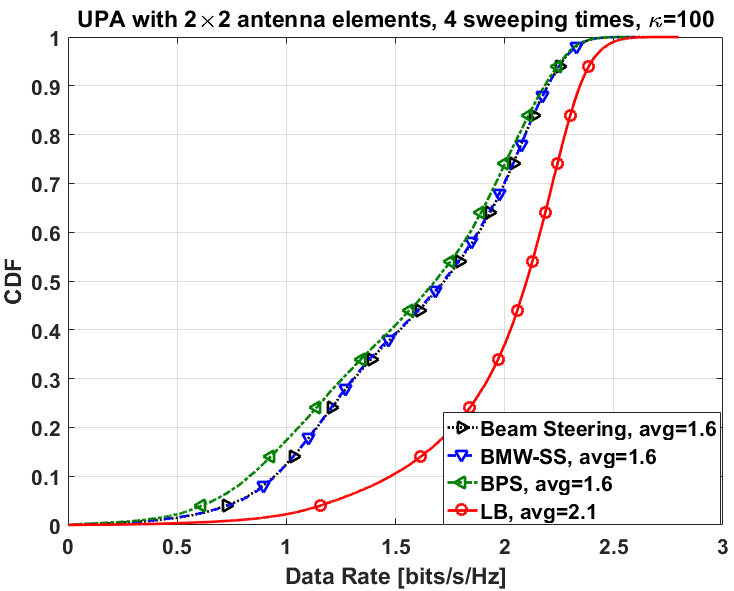}
}
\hfil
\subfloat[NLOS]{
\includegraphics[width=2in]{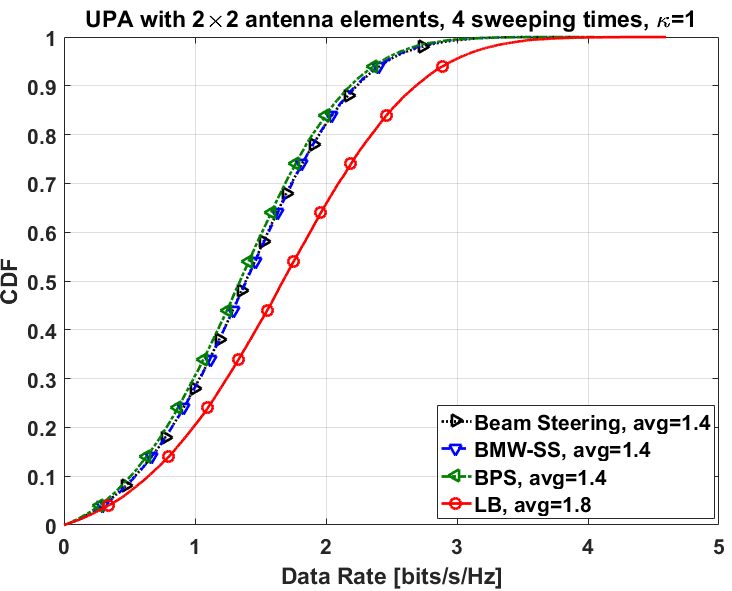}
}
\caption{CDF of Data rate for UPA with $2\times 2$ antenna elements}
\label{rate_2by2_4}
\end{figure}

The data rate can also be considered as a random variable denoted by $R(\mathbb{W})=log_2(1+\frac{P_{tot}}{N_0}G_B(\mathbb{W}))$ which depends on the distribution of the channel vectors and also the codebook design. We plot the CDF of $R(\mathbb{W})$ assuming $\frac{P_{tot}}{N_0}=5 dB$ and considering different codebooks including Beam-steering, BMW-SS, BPS and LB. The LB codebooks are designed by optimizing $J_{rate}$. In Fig. \ref{rate_8_4}, the CDF of $R(\mathbb{W})$ is shown for ULA with 8 antennas and 4 sweeping times, assuming $10^5$ different realizations of noise vector and channel vector for LOS and NLOS scenarios. The LB codebook provides $\overline{R}(\mathbb{W})=2.2$ and $\overline{R}(\mathbb{W})=1.8$ in LOS and NLOS scenarios, respectively, higher than that of other codebooks. It also provides better outage for every data rate threshold. For example, LB codebook provides $Pr(R(\mathbb{W})<1)=9\%$ in LOS scenario, while BPS, BMW-SS and beam-steering codebooks provide $25\%$, $28\%$ and $42\%$ data rate outage. Similar analysis is shown in Fig. \ref{rate_2by2_4}, for $2\times2$ UPA with 4 codewords. Higher average data rate and lower data rate outage are provided by using the LB codebook. 
\subsection{Quantized Phase Shifters}
\begin{figure}[!h]
\centering
\subfloat[LOS]{ 
\includegraphics[width=2in]{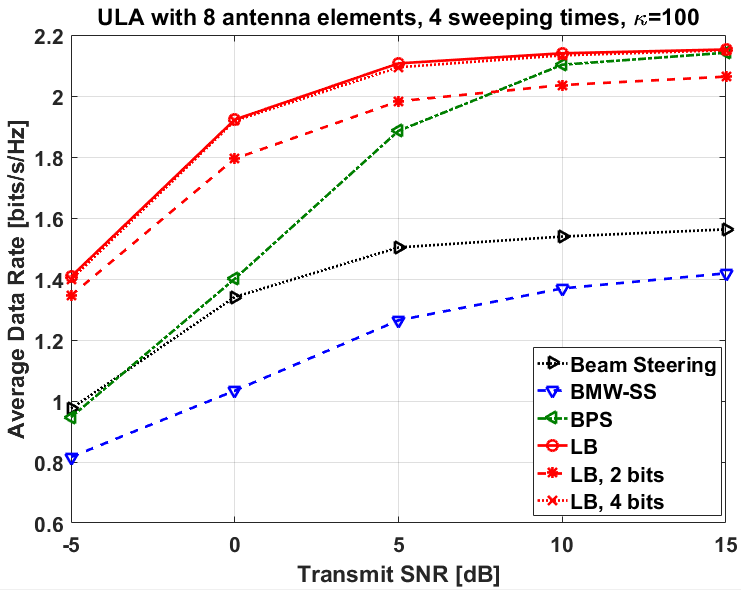}
}
\hfil
\subfloat[NLOS]{
\includegraphics[width=2in]{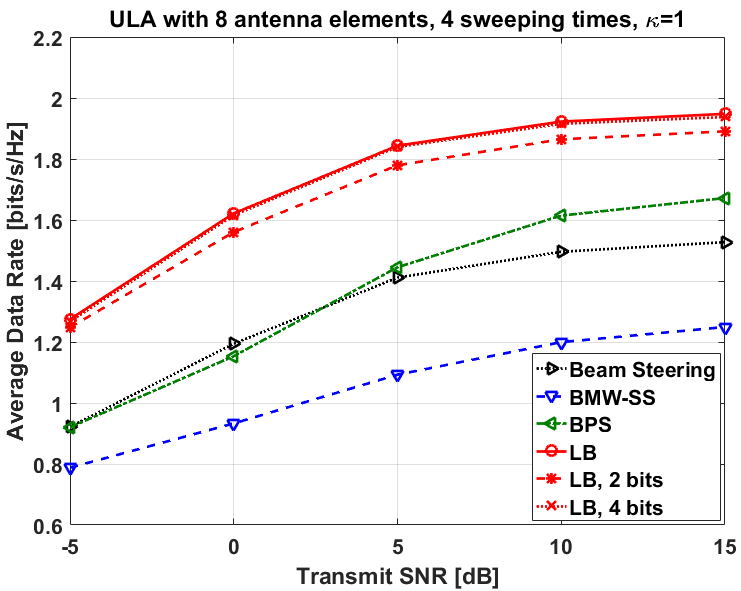}
}
\caption{Performance of quantized LB codebook with 4 sweeping times for 8 antenna ULA}
\label{quant_8_4}
\end{figure}

\begin{figure}[!h]
\centering
\subfloat[LOS]{ 
\includegraphics[width=2in]{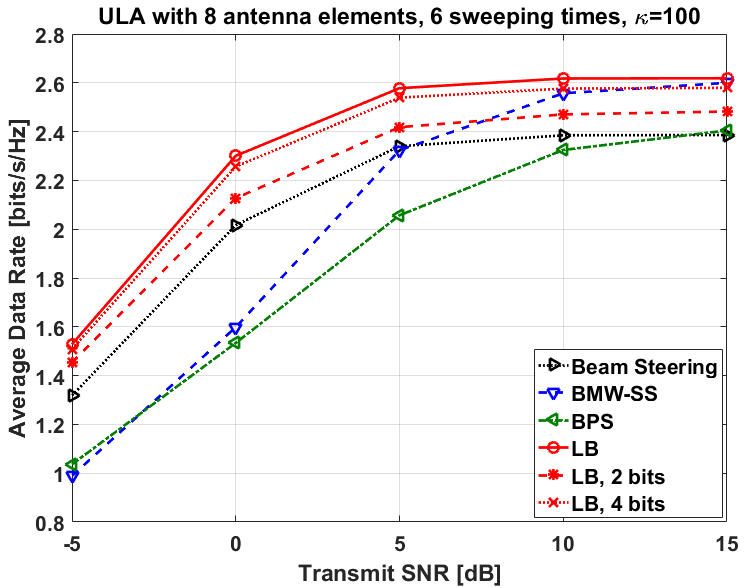}
}
\hfil
\subfloat[NLOS]{
\includegraphics[width=2in]{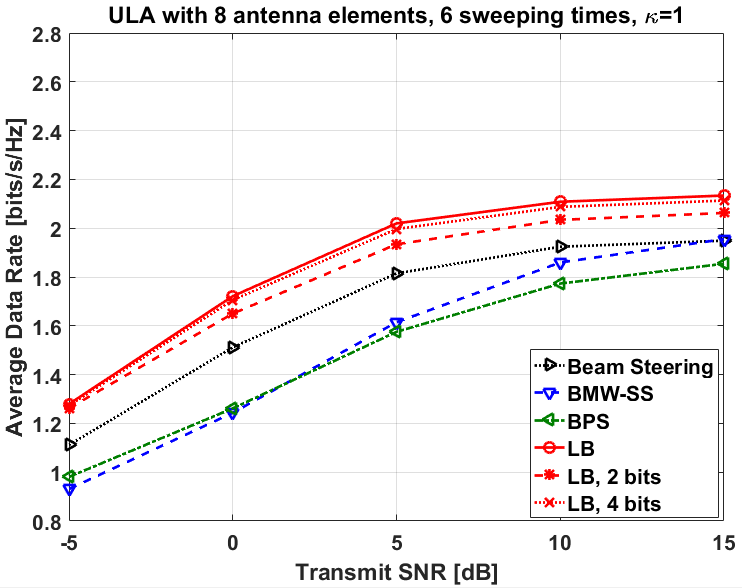}
}
\caption{Performance of quantized LB codebook with 6 sweeping times for 8 antenna ULA}
\label{quant_8_6}
\end{figure}

\begin{figure}[!h]
\centering
\subfloat[LOS]{ 
\includegraphics[width=2in]{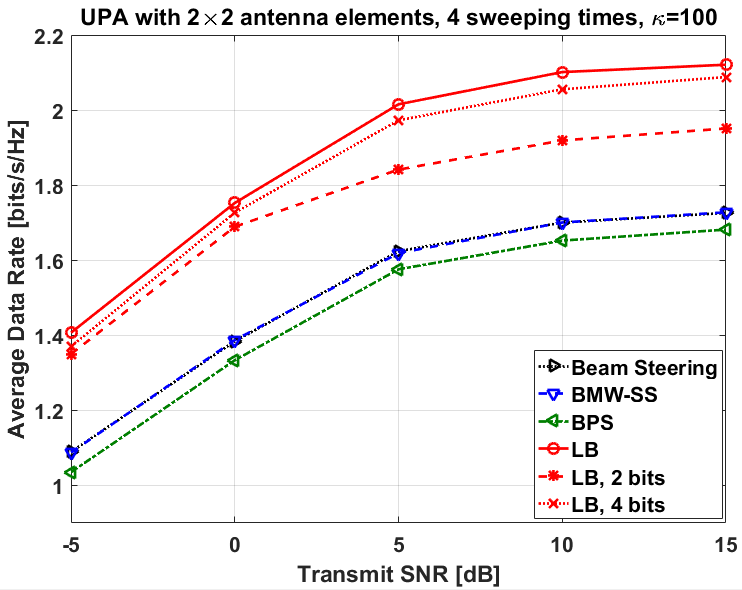}
}
\hfil
\subfloat[NLOS]{
\includegraphics[width=2in]{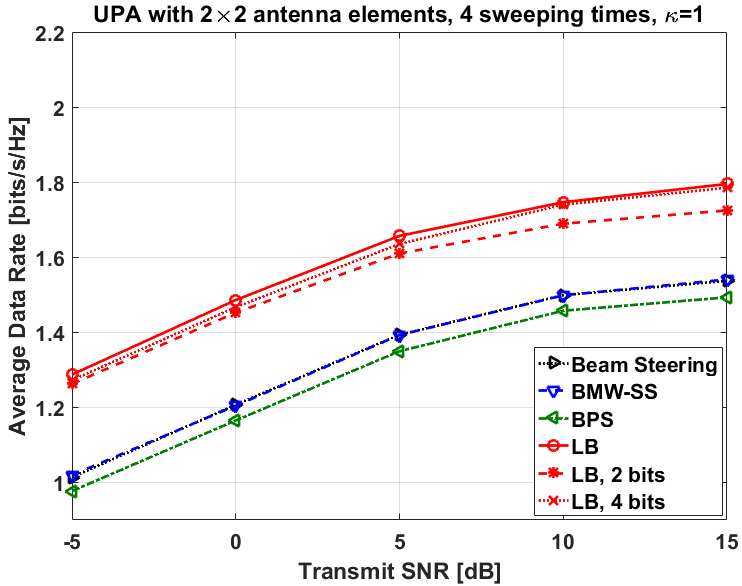}
}
\caption{Performance of quantized LB codebook with 4 sweeping times for $2\times2$ UPA}
\label{quant_2by2_4}
\end{figure}

 In this section, performance of the quantized codebooks designed with LB algorithm are analyzed. The average data rate, i.e., $\overline{R}(\mathbb{W})$, is considered and it is shown that the LB codebook with 2-bit phase shifters can outperform other codebooks with ideal phase shifters. In Figs. \ref{quant_8_4} and \ref{quant_8_6}, 8-antenna ULA with 4 and 6 sweeping times are considered, respectively. The codebook with 4-bit quantized phase shifters provides almost the same average data rate and using 2-bit phase shifters suffers from $\approx 0.1(5\%)$ loss in average data rate. Same analysis for $2\times2$ UPA with 4 codewords is provided in Fig. \ref{quant_2by2_4}. 
\vspace{-10pt}
\section{Conclusion}\label{conclusion}
In this work, we studied the codebook design for analog beamforming and proposed a new framework to design beamforming codebooks. The proposed method is inspired by generalized Lloyd algorithm and is able to optimize various performance metrics including, but not limited to, average beamforming gain, outage, and average data rate. In addition, unlike most of the work in the literature, which are limited to ULAs, the proposed algorithm can be applied to any array shapes with arbitrary phase offset vectors. We showed that our proposed codebook outperforms the codebooks in the literature with hierarchical designs; however, our framework can also be integrated with hierarchical designs by adjusting the objective function and the training channel vectors. We also considered quantized phase shifters and showed the superiority of our quantized codebooks when compared with the existing codebooks in the literature. In a nutshell, the proposed framework enables us to fully exploit analog beamforming; however, it can be further improved by making more diligent choices regarding training points, initialization, a stopping criterion, and number of iterations. The inclusion of an on-line adaption to modify the codebook each time a new subspace sample is observed can be possibly beneficial and interesting to analyze.    
\bibliographystyle{IEEEtran}
\bibliography{ref}

\addtolength{\textheight}{-12cm}   % This command serves to balance the column lengths
                                  % on the last page of the document manually. It shortens
                                  % the textheight of the last page by a suitable amount.
                                  % This command does not take effect until the next page
                                  % so it should come on the page before the last. Make
                                  % sure that you do not shorten the textheight too much.

%%%%%%%%%%%%%%%%%%%%%%%%%%%%%%%%%%%%%%%%%%%%%%%%%%%%%%%%%%%%%%%%%%%%%%%%%%%%%%%%

%%%%%%%%%%%%%%%%%%%%%%%%%%%%%%%%%%%%%%%%%%%%%%%%%%%%%%%%%%%%%%%%%%%%%%%%%%%%%%%%

%%%%%%%%%%%%%%%%%%%%%%%%%%%%%%%%%%%%%%%%%%%%%%%%%%%%%%%%%%%%%%%%%%%%%%%%%%%%%%%%
%\section*{APPENDIX}

\end{document}